\RequirePackage{rotating}
\documentclass[fleqn,usenatbib]{mnras}

\usepackage{newtxtext,newtxmath}

\usepackage[T1]{fontenc}
\usepackage{ae,aecompl}

\usepackage{graphicx}	
\usepackage{amsmath}	
\usepackage{amssymb}	
\usepackage{adjustbox} 
\usepackage{pdflscape} 
\usepackage{rotating}
\usepackage{afterpage}
\usepackage[graphicx]{realboxes}
\usepackage{varwidth}

\title[Voids in Evolving Dark Sector Cosmologies]{Cosmic Voids in Evolving Dark Sector Cosmologies: the High Redshift Universe}

\author[E. Adermann et al.]{
Eromanga Adermann,$^{1}$\thanks{E-mail: eromanga.adermann@sydney.edu.au (EA)}
Pascal J. Elahi,$^{2}$
Geraint F. Lewis$^{1}$
and Chris Power$^{2}$
\\
$^{1}$Sydney Institute for Astronomy, School of Physics, A28, The University of Sydney, NSW, 2006, Australia\\
$^{2}$International Centre for Radio Astronomy Research (ICRAR), The University of Western Australia, 35 Stirling Hwy,\\
Crawley, Western Australia 6009, Australia
}

\date{Accepted 2018 July 5. Received 2018 June 12; in original form 2018 February 22}

\pubyear{2018}

\begin{document}
\label{firstpage}
\pagerange{\pageref{firstpage}--\pageref{lastpage}}
\maketitle

\begin{abstract}
We compare the evolution of voids formed under the standard cosmological model and two alternative cosmological models. The two models are a quintessence model ($\phi$CDM) and a Coupled Dark Matter-Dark Energy (CDE) model, both of which have evolving and interacting dark sectors. From $N$-body adiabatic hydrodynamical simulations of these models, we measure the statistics and quantify the properties of voids over the redshift range $z=1.5-12$: these include their population size, volumes, shapes and average densities. We find that the latter property has potential as a probe of cosmology, particularly dark energy, as significant differences in average void densities exist between the alternative models and the standard model. We postulate that this signature arises from an increased evacuation rate of particles out of voids, or an earlier start to void evacuation, in the alternative models as a direct consequence of the dynamical scalar field, which also leads to greater void merger rates. Additionally, differences between the two alternative models are likely due to the drag force arising from dark sector coupling, acting on dark matter particles in our coupled model.  
\end{abstract}




\section{Introduction}
The Standard Model of Cosmology, $\Lambda$ Cold Dark Matter ($\Lambda$CDM), is the simplest and most successful cosmological model of the Universe. It asserts that the universe is spatially flat and dominated by a dark sector, comprised of non-baryonic dark matter and a dominant component of dark energy. 

The presence of dark matter is seen through observations of galaxy rotation dynamics and cluster dynamics \citep[e.g.][]{2006ApJ...648L.109C}. It is commonly assumed to be a non-relativistic (or `cold'), non-baryonic form of matter that interacts primarily via the gravitational force, and very likely played an important role in the formation of structure by driving the small-scale clustering of baryonic matter \citep[see][]{2012AnP...524..507F}. The exact properties, such as mass, of the dark matter particle(s) are unknown, although there are a number of theoretically-motivated candidates from high energy physics whose properties are consistent with observations, for example, the lightest supersymmetric particle (LSP), the Weakly Interacting Massive Particle (WIMP) and the Strongly Interacting Massive Particle \citep[SIMP;][]{2015PhRvL.115b1301H}. Dark energy was discovered through measurements of the redshift-luminosity distance relationship in Type Ia supernovae \citep{1998AJ....116.1009R, 1999ApJ...517..565P}, which revealed the late-time acceleration of cosmic expansion. It is considered a form of negative pressure vacuum energy driving the acceleration, and is characterised by the cosmological constant $\Lambda$ and the equation of state $w=-1$. 

The $\Lambda$CDM model has had enormous success in explaining large-scale observations, such as the Cosmic Microwave Background (CMB) anisotropies \citep{1999Sci...284.1481B, 2013ApJS..208...20B, 2014A&A...571A..16P, 2016A&A...594A..16P}, Baryonic Acoustic Oscillations \citep{2011MNRAS.418.1707B, 2012MNRAS.423.3430B, 2014MNRAS.441...24A}, features in the large-scale structure \citep[e.g.][]{2012MNRAS.423.3430B}, weak gravitational lensing by the large-scale structure \citep[e.g.][]{2013MNRAS.430.2200K} and galaxy clustering \citep[e.g.][]{2017MNRAS.470.2617A}. As such, it has become the standard model for comprehending the Universe on large-scales. 

Although $\Lambda$CDM is well-supported by observations, tensions still exist between some observations and its predictions. For example, $\Lambda$CDM overestimates the number of satellite galaxies around galaxies similar to the Milky Way \citep[e.g.][]{1999ApJ...522...82K, 1999ApJ...524L..19M}. While this tension could be due to insufficient modelling of feedback processes and their effects (e.g. small sub-haloes may not be able to host satellite galaxies due to removal of gas from feedback processes, as suggested by \citealt{2000ApJ...539..517B, 2002MNRAS.333..177B, 2011MNRAS.415..257N, 2012ASPC..453..305N, 2016MNRAS.457.1931S}), it could also indicate the need for modifications to $\Lambda$CDM, perhaps in the form of warm dark matter \citep[for details, see][]{2012MNRAS.424..684S, 2013PASA...30...53P, 2014MNRAS.439..300L, 2014MNRAS.444.2333E}, although some studies have shown that warm dark matter does not eliminate small-scale inconsistencies \citep[e.g.][]{2014MNRAS.441L...6S}. $\Lambda$CDM also cannot explain or predict the strong alignment of satellite galaxies observed in the local group (e.g. the Vast Polar Structure; \citealt{2012MNRAS.423.1109P, 2018arXiv180202579P} and the Plane of Satellites; \citealt{2013Natur.493...62I, 2013ApJ...766..120C} or beyond (e.g. alignments in SDSS data; \citealt{2006MNRAS.369.1293Y, 2013ApJ...768...20L,2014Natur.511..563I}), indicating a further need for extensions or alterations to the model.  

The $\Lambda$CDM model also has some theoretical shortcomings. One such example is the so-called `cosmological constant problem', where the prediction for the value of $\Lambda$ from quantum field theory (assuming that dark energy arises from the zero-point energy of a fundamental quantum field) is stunningly inconsistent with the value of $\Lambda$ derived from observations and general relativity, which could be a sign of a  fundamental misrepresentation of dark energy by the standard model \citep[for a review, see][]{1989RvMP...61....1W, 2012CRPhy..13..566M}. Another issue is that the density of dark energy is similar in value to the density of matter today, despite their independent evolution through cosmic time and the very small period in which they should be comparable \citep[for a review, see][]{2002CQGra..19.3435S}. This unlikely coincidence may point to some inter-dependence within the dark sector that is not accounted for in $\Lambda$CDM. Perhaps most frustrating is the lack of theoretical underpinning for dark energy and dark matter -- though we can characterise the dark sector with cosmological parameters, its nature remains a mystery. 

To address the shortcomings of the standard model, a number of alternative cosmological models have been proposed. For example, the coincidence problem has led to the proposal of models with a time-dependent dark energy density and equation of state. These include quintessence models, which feature a dynamical scalar field that drives the accelerated universal expansion instead of a cosmological constant \citep[see][]{PhysRevD.35.2339, 1988NuPhB.302..668W, 1988PhRvD..37.3406R, 2013CQGra..30u4003T}. For specific types of potentials, the scalar field density remains close to the matter density throughout most of its evolution, which alleviates the cosmic coincidence problem. Another example of a class of dynamical dark energy models are the $f(R)$ gravity models \citep[for a review, see][]{2010LNP...800...99T}, which were proposed to explain dark energy through modifications to gravity on large-scales \citep[e.g.][]{2002IJMPD..11..483C, 2004PhRvD..70d3528C}. It is worth mentioning that these models have not been ruled out by the recent confirmation that gravitational waves travel at light speed, while others, such as Galileons, quartic and quintic Horndeski theories, and some beyond-Horndeski models have \citep[see, for example,][]{2017arXiv171005877C, PhysRevLett.119.251304, 2017arXiv171006394B}.  

To constrain the possible models that could explain observations of our Universe, cosmologists have utilised the properties and statistics of the large-scale structure, in particular the overdense sub-structures (e.g. the halo mass function \citealt{2008ApJ...687....7S}) and galaxy cluster gas properties \citep{2010MNRAS.403.1684B, 2014MNRAS.439.2958C}. However, in recent times, cosmic voids (underdense regions occurring within the LSS) have gained momentum as a probe of cosmology, as they are only mildly non-linear structures (due to their low density), which makes them potentially more sensitive probes than overdense structures like haloes and filaments. The properties and growth of overdense structures are heavily influenced by complex baryonic physics and gravitational interactions, meaning that cosmological signatures may be difficult to distinguish from, or even be masked by, non-cosmological effects. In contrast, cosmic voids are relatively simple environments that are much less affected by baryonic physics, meaning that any cosmological signatures in their properties are easier to identify. As such, they have been established as useful probes of dark energy \citep[e.g.][]{2012MNRAS.426..440B, 2015PhRvD..92h3531P}, modified gravity \citep[e.g.][]{2017PhRvD..95b4018V} and alternative cosmological models \citep[e.g.][]{2015JCAP...11..018M, 2015MNRAS.446L...1S, 2016MNRAS.455.3075P, 2017MNRAS.468...59W, 2017MNRAS.468.3381A, 2017PhRvD..96h3506A}. Voids have also been used in studies and measurements of the integrated Sachs-Wolfe effect signal \citep[e.g.][]{2014ApJ...786..110C, 2015MNRAS.454.2804G, 2017arXiv170108583K}, baryonic acoustic oscillations (in void clustering; e.g. \citealt{2016PhRvL.116q1301K, 2016MNRAS.459.4020L}), and weak lensing studies \citep[e.g.][]{2016MNRAS.455.3367G, 2017MNRAS.465..746S}.  

This paper will make use of voids as probes of cosmology, extending the results presented in \citet{2017MNRAS.468.3381A} to the high redshift universe, in order to identify discrepancies between cosmologies at all times, including in the dark matter dominated era. Expanding to high redshift allows us to potentially differentiate between the effects of normalisation (those which are always present and continuously evolving, and those which only appear in the dark energy dominated era) and cosmology. We focus on the properties of voids and their evolution from z=1.5-12, a redshift range seldom explored in studies of voids. In Sections \ref{Models}, \ref{Simulations} and \ref{Method}, we present a summary of the models examined in this paper, details of the simulations of these models, and our approach to finding voids in the simulations. In Section \ref{Results}, we present our results, followed by our interpretation and discussion in Section \ref{Discussion}. In Section \ref{Conclusions}, we summarise the findings and implications of the results in this paper. 

\section{Evolving Dark Sector Models}\label{Models}
We compare predictions from the standard model to predictions from two non-standard, evolving dark sector models: an uncoupled quintessence model ($\phi$CDM) and a coupled dark energy-dark matter model, which we shall refer to as Coupled Dark Energy (CDE) throughout this paper. The key difference between the non-standard models and $\Lambda$CDM is the nature of dark energy. Within the non-standard models, a time-dependent scalar field, $\phi$, is responsible for dark energy, rather than the cosmological constant, $\Lambda$. The CDE model is further distinguished from both $\phi$CDM and $\Lambda$CDM, in that its scalar field is coupled to the dark matter field. 

We choose a Ratra-Peebles potential \citep{1988PhRvD..37.3406R} for both our non-standard models, given by: 
\begin{equation}
V(\phi)=V_{0}\phi^{-\alpha}
\end{equation} 
where $V_{0}$ and $\alpha$ are constants. 

The consequence of including the scalar field in the uncoupled quintessence model and the coupled dark energy model is that the dark energy density evolves with time, which affects the expansion history of the universe. However, the Lagrangian for the scalar field of the coupled dark energy model contains a coupling term, which allows dark matter particles to decay into the scalar field. An additional consequence of the coupling is the presence of a `frictional' force acting on dark matter particles, which affects the evolution of density perturbation amplitudes. For a more detailed discussion of these models, see \citet{2015MNRAS.452.1341E} and \citet{2017MNRAS.468.3381A}. 

\section{Simulations}\label{Simulations}
We produced $N$-body simulations of the two models of interest, $\phi$CDM and CDE, and a reference $\Lambda$CDM simulation to compare to the alternative models. We choose the coupling parameter, $\beta_{o}$, in the CDE simulation to be 0.05. The choice allows us to maximise any differences between the model and the reference $\Lambda$CDM model, while still remaining within the range of allowed coupling \citep[see][]{2012PhRvD..86j3507P, 2013JCAP...11..022X}. 

Our simulations were run using DARK-GADGET, a modified version of P-GADGET-2, which is itself a modified version of GADGET-2 \citep[for more details, see][]{2005MNRAS.364.1105S, 2014MNRAS.439.2943C}. Our implementation follows that of \cite{2014MNRAS.439.2943C}. We include a separate gravity tree to account for the long range forces arising from the scalar field, and an evolving dark matter $N$-body particle mass for CDE, arising from the decay of the dark matter density. The simulations each had a mass resolution of $m_{dm}(m_{gas}) = 6.9(13) \times 10^{10} h^{-1} M_{\sun}$ at $z=0$, high enough for void identification. Each simulation was contained in a box of size 500 $h^{-1}$Mpc in length and consisted of $2 \times 512^{3}$ particles (dark matter and gas). The linear power spectrum and the growth factor were calculated using the publicly available Boltzmann code CMBEASY \citep{2005JCAP...10..011D} and first-order Newtonian perturbation equations. For plots of the growth factor evolution, the Hubble constant evolution and the non-linear power spectrum in each of the simulations, please refer to Section 3 of \cite{2017MNRAS.468.3381A}. 

The initial conditions for the simulations were produced using a modified version of the publicly available N-GENIC, by perturbing particles in a Cartesian grid with the first order Zel'dovich approximation. The modified N-GENIC code uses the growth factors $f = d ln D(a)/d ln a$ (which were calculated by CMBEASY) to determine the particle displacements in the alternative cosmologies. We chose to match the cosmological parameters $h$, $\omega_{m}$, $\omega_{b}$ and $\sigma_{8}$ among the simulations at $z=0$, rather than $z=z_{CMB}$ (redshift of the Cosmic Microwave Background). The values were chosen to be consistent with the $z=0$ $\Lambda$CDM Planck data \citep{2014A&A...571A..16P, 2016A&A...594A..13P}, with ($h$, $\Omega_{m}$, $\Omega_{b}$, $\sigma_{8}$) = (0.67, 0.3175, 0.049, 0.83) for all three simulations. Matching the parameters at $z=0$ results in the alternative models having a different expansion history in the early universe compared to $\Lambda$CDM. Each simulation begins at $z=100$, with the same initial density perturbation phases, resulting in underdense regions forming in the same approximate locations in each simulation. This leaves any differences among the models to manifest in the density profiles of the underdense regions (according to their individual power spectra), and minimises differences arising from cosmic variance. 

Although the simulations do not take into account star formation or feedback physics, recent studies have shown that this should not significantly affect the void population. \cite{2017MNRAS.470.4434P} demonstrated that large underdensities remain underdense, regardless of whether full baryonic/hydrodynamic physics is accounted for. 

\section{Void Finding}\label{Method}
Following \citet{2017MNRAS.468.3381A}, we identified voids in the cold dark matter particle distribution in the simulations by calculating the Hessian matrix across the density field, which is given by: 
\begin{equation}
H_{\alpha \beta}(\textbf{x})=\frac{\partial^{2}\rho(\textbf{x})}{\partial x_{\alpha} \partial x_{\beta}}. 
\end{equation}

The Hessian matrix can be used to characterise the curvature of the density field, $\rho(\textbf{x})$ at position $\textbf{x}$. If all three eigenvalues are negative, the local density field exhibits a minimum along all directions, and is thus void-like. Various combinations of positive and negative eigenvalues characterise sheet-like, filament-like and knot-like regions within the density field. 

To calculate the Hessian matrix and its eigenvalues across an entire simulation box, we first divided it up into a grid of $500 \times 500 \times 500$ cubic cells, each $1$ $h^{-1}$Mpc in length. Particle densities were calculated using a smooth particle hydrodynamics kernel, and then were assigned to the nearest grid cell, and used to determine an average density for each cell. The densities were then convolved with a Gaussian kernel, with $\sigma = 3$ $h^{-1}$Mpc, in order to remove small-scale noise in the density field. We then calculated the Hessian matrix for each cell, and used its eigenvalues to identify the cells that were void-like. We grouped the neighbouring void-like cells together with a friends-of-friends algorithm, with the requirement that neighbouring cells could only be linked if both are void-like cells, or the second is a sheet-like cell. Each void group thus included a boundary layer of sheet-like cells of thickness $1$ $h^{-1}$Mpc. Void groups were only classified as voids if their total volume consisted of at least two void-like cells linked together (with a corresponding volume of $2$ $h^{-3}$Mpc$^{3}$), as smaller groups consisting of one void-like cell may be spurious voids, and the resolution of our grid does not allow us to distinguish between these voids and genuine voids with volume $1$ $h^{-3}$Mpc$^{3}$.  

Recent structure finder comparison projects \citep[e.g.][]{2018MNRAS.473.1195L} have revealed that there is good agreement between most void finding methods, especially in the shape of the void density profiles. We note that the Hessian-based methods were associated with smaller volume filling fractions than other methods, which is consistent with our results. However, the void filling fraction within our simulations is still smaller than what was calculated using other Hessian-based methods, potentially because of the difference in the eigenvalue thresholds used. While other methods set their threshold based on visual comparisons, we set ours based on a physical definition of voids as density minima. 

We also suggest that other, non-Hessian methods tend to identify much larger underdense regions than Hessian-based methods, because Hessian methods rely on density changes in space, allowing the identification of individual density minima within these larger underdensities. Although a small fraction of density minima identified in this way could be ``voids-in-clouds'' (approximately $0.3\%$ of voids we identified have an average density greater than the average simulation density at $z=0$), we argue that this should not be problematic, as long as there is consistency in the methods used to define and identify voids when comparing predictions to observations. More information on how our void-finding method compares to other methods can be found in \cite{2017MNRAS.468.3381A} and \cite{2018MNRAS.473.1195L}. 

\section{Results}\label{Results} 
We present the population size, volume distribution, shape distribution and average density distribution of voids occurring across the high redshift universe. Our analysis spans the redshift range from $z=1.5$ to $z=12$. 

\subsection{Population Size}\label{Popsize}
\begin{figure}
\includegraphics[trim={0.4cm 0.1cm 0.9cm 1.4cm}, clip, width=\linewidth, keepaspectratio]{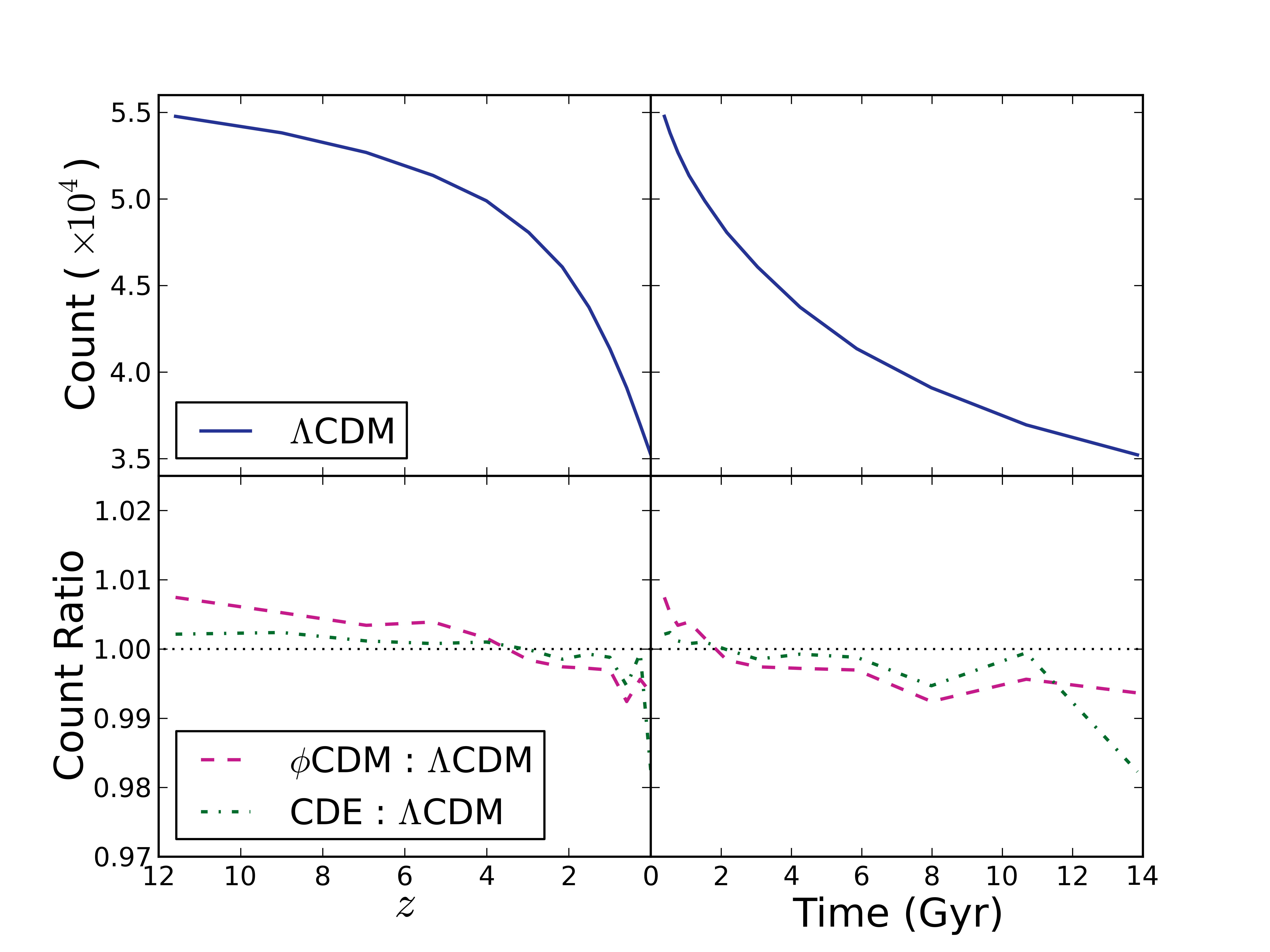}
\caption{Left: The evolution of the total number of voids found in the $\Lambda$CDM model across the redshift range $z=0-12$ (top panel), and the ratio of the population size in the $\phi$CDM (dashed magenta) and CDE (dotdashed green) models to the $\Lambda$CDM model (bottom panel). A ratio of one is indicated by the black dotted line. Right: The evolution of the total number of voids in each model as a function of the age of the Universe, calculated using the same cosmological parameters used in our simulations.} 
\label{voidcounts} 
\end{figure}

The evolution of the void population size for $\Lambda$CDM is shown in Figure~\ref{voidcounts}, along with the ratios of the $\phi$CDM and CDE population size to the $\Lambda$CDM population size. The number of voids decline with time for all models, and is generally very similar between the models at all redshifts, with all ratios staying close to unity. However, there are small differences between the models worth noting. 

Specifically, the $\phi$CDM and the CDE models have greater void populations than the $\Lambda$CDM model at high redshift ($z\approx4-12$), with $\phi$CDM having the largest population. At $z\approx4$, the $\Lambda$CDM void population begins to overtake both $\phi$CDM and CDE, and continues to be most populated until $z=0$. In the range $z\approx0.3-0.6$, the population size ratios between the alternative models and $\Lambda$CDM increase slightly, before decreasing again towards $z=0$. At $z\approx0.3$, the population of CDE suddenly declines and the model becomes less populated than $\phi$CDM. 

\subsection{Volumes}
Following \cite{2017MNRAS.468.3381A}, we use a probability density function (PDF) to quantify the shape of the void volume distribution across the high redshift universe. The PDF is defined so that the probability of finding a void with a specific volume $V$ is given by the integral from $V$ to $V+dV$ of the PDF. The number of voids of a given volume that can be expected to exist in a given region can be determined from the PDF by multiplying the probability of finding a void with that volume by the number of voids inside the region. 
The probability density function is given by: 
\begin{equation}\label{volPDF}
f(V) = \frac{1}{V_{0} \Gamma(1-\alpha)}(V/V_{0})^{-\alpha} \mathrm{exp}(-V/V_{0}), 
\end{equation}
where $V$ is the void volume in $h^{-1}$Mpc, defined as the number of $1$ $h^{-1}$Mpc void-like cells comprising the void (excluding sheet-like cells). The parameter $\alpha$ is the slope of the power-law, $V_{0}$ is the characteristic volume that determines the position of the turnover, and $\Gamma(1-\alpha)$ is the gamma function evaluated at $1-\alpha$. 
The best fit values for the parameters $\alpha$ and $V_{0}$ were determined by fitting Equation \ref{volPDF} to the set of void volumes at each redshift in all simulations, and are shown in Table \ref{volumefits}. The fitting was performed using a Markov Chain Monte Carlo (MCMC) algorithm with the Python emcee library \citep{emcee}. 


\begin{figure*}
\includegraphics[width=\textwidth, keepaspectratio]{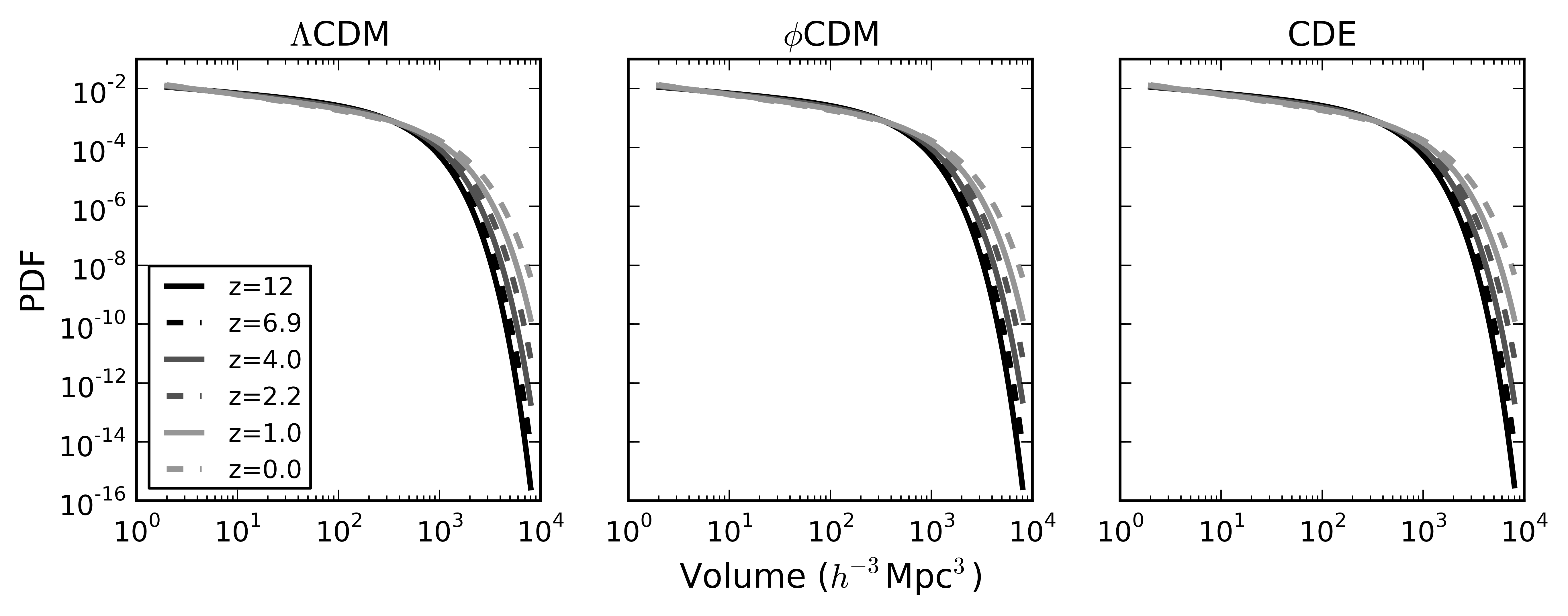}
\caption{The volume PDF distributions for $\Lambda$CDM, $\phi$CDM and CDE at multiple redshifts.}
\label{evol2} 
\end{figure*}

The PDF distributions in the $\Lambda$CDM simulation at a select number of redshifts is presented in Figure~\ref{evol2}. It is clear from the figure that the lower the redshift, the greater the probability of finding large voids ($V\gtrsim10^{3}$ $h^{-3}$Mpc$^{3}$). This trend is exhibited across the entire set of snapshots we analysed. However, in the volume range $V\approx4-300$ $h^{-3}$Mpc$^{3}$, a different trend emerges; the lower the redshift, the lower the probability of finding these voids. The PDF distributions for different $z$ intersect with each other at $V\approx300-310$ $h^{-3}$Mpc$^{3}$, because the higher redshift curves experience a faster drop off than the lower redshift curves (due to a combination of the slightly shallower power-law slope and the exponential drop off at lower volumes for higher redshift curves). At the lowest volumes ($2$ $h^{-3}$Mpc$^{3}$ $\leq$ $V\lesssim4$ $h^{-3}$Mpc$^{3}$) the PDFs show the same trend as at high volumes; the smallest voids occur more frequently at low redshift than at high redshift. Although the rate of growth and coalescence of these smallest voids may not be numerically converged, the trends observed are likely physical, and supported by the large number of small voids. The power-laws at lower redshifts are shallower than at higher redshifts, resulting in another intersection between the PDF distributions at approximately $4-5$ $h^{-3}$Mpc$^{3}$.  However, it must be noted that the exact locations of these intersection points are not well-constrained, as it occurs at low volumes very close to the noise limit. 

The PDF distributions across multiple redshifts for $\phi$CDM and CDE show similar trends. The intersection points between the PDF distributions for different $z$ also occurs at $V\approx300-310$ $h^{-3}$Mpc$^{3}$ and $V\approx 4 h^{-3}$Mpc$^{3}$ for $\phi$CDM and CDE. However, the range included between these crossover points is slightly smaller for $\Lambda$CDM than for the alternative models. Interestingly, the intersection between the $z=9$ and $z=12$ distributions in CDE appears to occur at less than $2$ $h^{-3}$Mpc$^{3}$, outside the range probed by our void finding method. Thus, in the CDE simulation, there are a slightly greater number of the smallest voids at $z=9$ than at $z=12$, which is not the case for the other two models. 

\begin{figure*}
\includegraphics[trim={3.0cm 0.5cm 3.5cm 1.3cm}, clip, width=\textwidth]{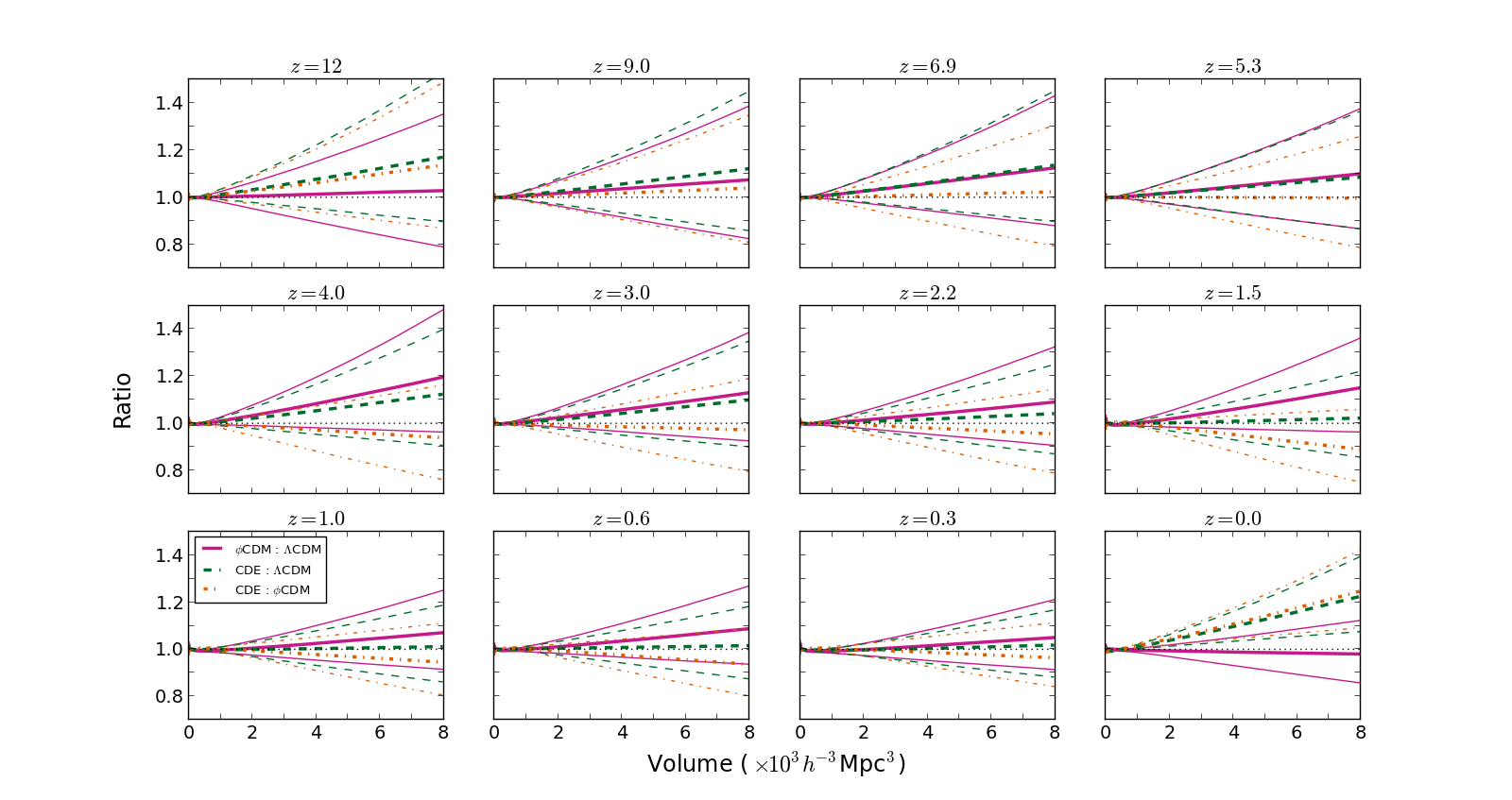}
\caption{Bootstrapped ratios for the volume PDFs between each pair of the three models, from $z=12-0$. We show the median ratio, along with the 16$^{\mathrm{th}}$ and 84$^{\mathrm{th}}$ percentile ratios, at each redshift. The median ratios are represented by the thickest lines, while the 16$^{\mathrm{th}}$ and 84$^{\mathrm{th}}$ percentile ratios are indicated by the thinner lines. The ratios between the $\phi$CDM PDF and the $\Lambda$CDM PDF are indicated by the solid magenta lines, while the dashed green lines and the dotdashed orange lines represent the ratios between the CDE and $\Lambda$CDM PDFs and between the CDE and $\phi$CDM PDFs respectively.}
\label{volPDFs}
\end{figure*}

Comparisons of the volume PDF between the models for $z=1.5-12$ are presented in Figure~\ref{volPDFs} (along with the $z=0-1$ comparisons for reference), where we show the ratio between each pair of cosmologies. To obtain these ratios, we performed a bootstrapping analysis on the distribution of parameter values that were sampled during the MCMC fitting process. We took a subsample of 5000 values for each of the two parameters, $\alpha$ and $V_{0}$, and calculated the ratios between the corresponding volume PDFs they defined. We were thus able to produce a distribution of ratios for each pair of models. We show the median ratios between each pair of cosmologies in Figure~\ref{volPDFs}\footnote{We show the ratios of the bin-independent fits, rather than ratios of the raw volume data which would be bin-dependent.}, along with the 16$^{\mathrm{th}}$ and 84$^{\mathrm{th}}$ percentile ratios, which represent the $1\sigma$ uncertainty on the ratios obtained, as the distributions are not necessarily Gaussian. These ratio plots show that across all redshifts studied, the ratios between the PDFs for each pair of cosmologies are consistent with a ratio of unity to within $1\sigma$ uncertainty ($1\sigma$ uncertainty range overlaps with a ratio of $1$). We observe the same increasing spread in the ratio distributions with volume for the high redshift universe, as we saw in the low redshift universe, because the shape of the PDFs at high volume are not as well-constrained as at low volume due to lower numbers of large voids. The median ratios also tend to increase with volume, indicating that the disparity between cosmologies is greater in the large void population than the small void population.  

At very early times, the $\Lambda$CDM PDF is closer to $\phi$CDM than CDE. Specifically, CDE predicts more large voids $V > 1000$ $h^{-3}$Mpc$^{3}$ than the other two models. From $z=9.0 - 5.3$, CDE converges towards $\Lambda$CDM. At the same time, $\phi$CDM shows greater discrepancy from the standard model, suggesting faster production of large voids within $\phi$CDM. By $z=4.0$, $\phi$CDM contains the largest number of voids with $V > 1000$ $h^{-3}$Mpc$^{3}$, and $\Lambda$CDM the least. The disparity between $\phi$CDM and $\Lambda$CDM decreases from $z=3.0-2.2$, before increasing again by $z=1.5$. Despite these changes, it continues to have the largest number of voids in the range $z=3.0-1.5$. The CDE PDF gradually approaches the $\Lambda$CDM PDF from $z=12-1.5$, until their median ratio is nearly unity at $z=1.5$. However, we note that these systematic differences are affected by the absence of large-scale power. Hence, the evolution described here may not accurately represent the large void populations, although each simulation should be affected in the same way. 

Although not shown, the results for smaller voids ($V < 1000$ $h^{-3}$Mpc$^{3}$) also show no significant deviation from each other by more than $1\sigma$, but there are small discrepancies among the cosmologies. There is no clear trend for the smallest voids, with each model containing the greatest number of these voids at different redshifts. However, small voids are close to our noise limit, which may hide differences among the models. Across the redshift range $z=1.5-12$, $\Lambda$CDM contains the greatest number of mid-range voids ($V \approx 100$ $h^{-3}$Mpc$^{3}$; specific volume range varies with redshift). 

\begin{figure*}
\includegraphics[width=\textwidth]{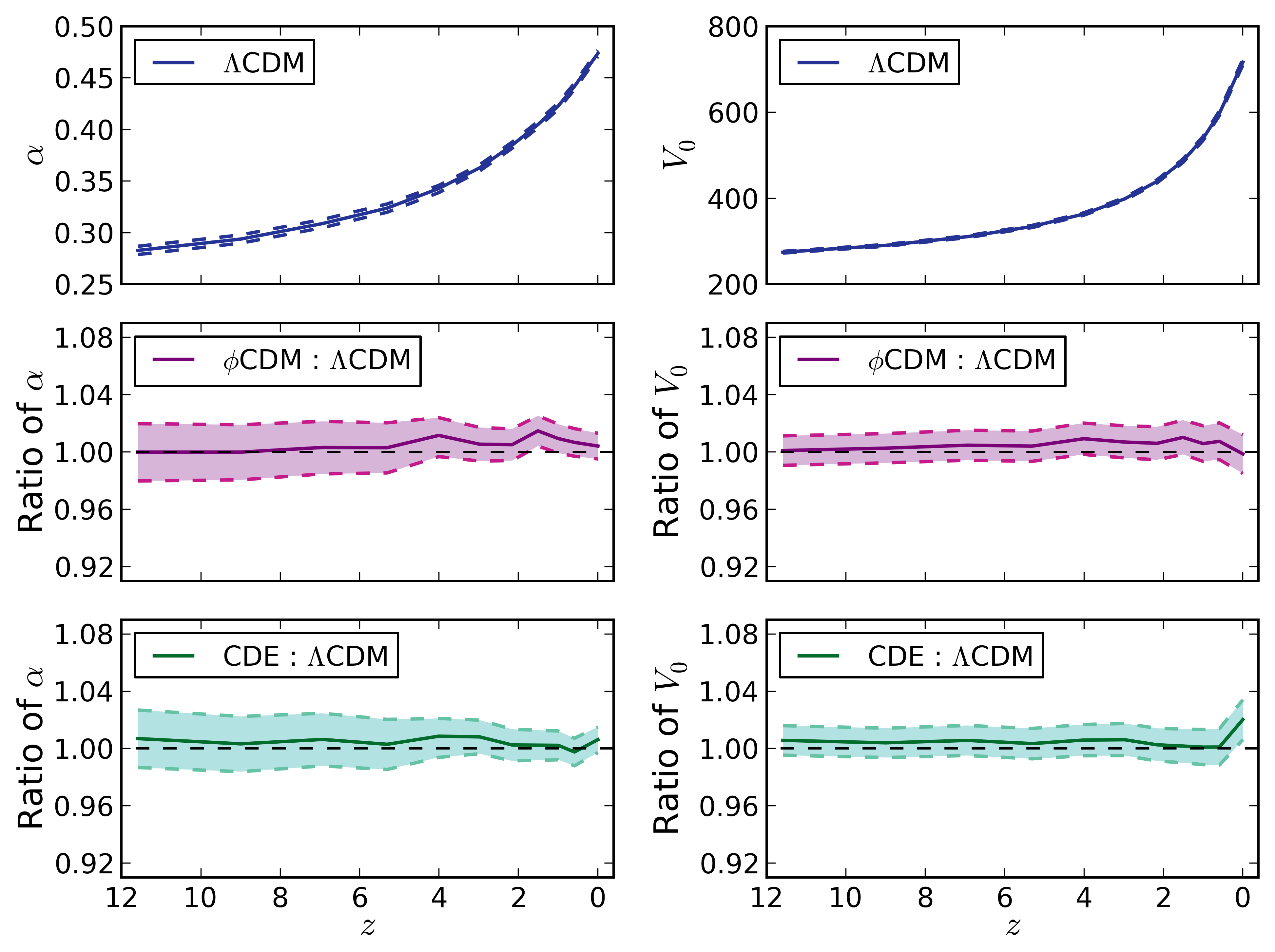}
\caption{The evolution of the parameters $\alpha$ and $V_{0}$ with redshift for the three models. The parameters characterising the void volume PDF in the $\Lambda$CDM model are displayed in the first row, with $1 \sigma$ uncertainties indicated by the dashed lines. The ratio of $\alpha$ and $V_{0}$ in the $\phi$CDM and CDE models to the corresponding $\alpha$ and $V_{0}$ in the $\Lambda$CDM model are shown in the second and third rows respectively.  The $1 \sigma$ spread on the ratios are indicated by the coloured regions. A ratio of unity is indicated by the black dashed lines.}
\label{volPDFimage}
\end{figure*}

In Figure~\ref{volPDFimage}, we present the evolution of the volume PDF parameters $\alpha$ and $V_{0}$. The ratios of the parameters in the $\phi$CDM and CDE models to the $\Lambda$CDM parameters are also presented. The uncertainties displayed were calculated by adding the parameter uncertainties in quadrature, while accounting for variation in the upper and lower uncertainties. It is clear from Figure~\ref{volPDFimage} that $\alpha$ and $V_{0}$ increase with decreasing redshift across all models, indicating that the power law steepness and the exponential cut-off point increases with decreasing redshift. Additionally, the rate of increase for $V_{0}$ increases as the models evolve (i.e. $d^{2}V_{0}/dt^{2} > 0$). Although this rate of increase is generally quite similar across the models, there is one difference; from $z=0-0.3$, the parameter $V_{0}$ in the CDE model increases much more quickly than in either of the other models. The parameter values are otherwise consistent among the models to within a $1\sigma$ uncertainty across all redshifts (uncertainty regions overlap a ratio of $1$). These trends are also clear from Table~\ref{volumefits}, which lists the best fit values of $\alpha$ and $V_{0}$ along with their $1\sigma$ uncertainties. \footnote{For reference, the best fit values from $z=0-1$ are also presented, which covers a total of $\sim 8$ Gyr in time.} 

\begin{table*}
\centering
\caption{The best fit values (with $1\sigma$ uncertainties) for the parameters $\alpha$ and $V_{0}$. The values for the $z=0.0-1.0$ are also included for the sake of comparison.}
\begin{tabular}{c c c c c c c c c}
    \hline
    &\multicolumn{2}{c}{$\Lambda$CDM} & &\multicolumn{2}{c}{$\phi$CDM} && \multicolumn{2}{c}{CDE}\\
    \cline{2-3}
    \cline{5-6}
    \cline{8-9} 
  	$z$ & $\alpha$ & $V_{0}$ & &$\alpha$ & $V_{0}$ & &$\alpha$ & $V_{0}$ \\
    \hline
    \hline
 	12 & $0.283\pm0.004$ & $275.0\pm2.0$ & &$0.283\pm0.004$ & $275.3\pm2.0$& &$0.285\pm0.004$ & $276.6\pm2.0$\\
  	9.0 & $0.294\pm0.004$& $290.9\pm2.1$& &$0.294\pm0.004$& $291.7\pm2.1$&& $0.295\pm0.004$& $292.1\pm2.1$\\
    6.9 & $0.309\pm0.004$& $311.2\pm2.3$& & $0.310\pm0.004$& $312.7\pm2.3$& &$0.311\pm0.004$ &$313.0\pm2.3$\\
    5.3 &$0.324\pm0.004$& $334.9\pm2.5$& &$0.325\pm0.004$ &$336.3\pm2.5$ & &$0.325\pm0.004$ &$336.1\pm2.5$\\
    4.0 &$0.343^{+0.003}_{-0.004}$& $363.8\pm2.8$& & $0.347\pm0.003$& $367.2\pm2.8$& & $0.346\pm0.003$ & $366.0\pm2.8$\\
    3.0 &$0.363\pm0.003$& $399.0^{+3.2}_{-3.1}$& &$0.365\pm0.003$ & $401.8^{+3.2}_{-3.1}$& &$0.366\pm0.003$ &$401.5^{+3.2}_{-3.1}$\\
    2.2 &$0.385\pm0.003$& $440.0^{+3.6}_{-3.5}$& &$0.387\pm0.003$ & $442.7\pm3.6$& &$0.386\pm0.003$ &$441.2^{+3.6}_{-3.5}$\\
    1.5 &$0.405\pm0.003$&$487.6^{+4.1}_{-4.0}$ & & $0.411\pm0.003$& $492.6^{+4.2}_{-4.1}$& &$0.406\pm0.003$ &$488.5^{+4.1}_{-4.0}$\\
    \hline
    1.0 &$0.423\pm0.003$&$540.0\pm4.7$& &$0.427\pm0.003$&$543.2^{+4.8}_{-4.7}$& &$0.424\pm0.003$&$540.6^{+4.7}_{-4.6}$\\
    0.6 &$0.442\pm0.003$&$598.3^{+5.4}_{-5.3}$& &$0.445\pm0.003$&$602.8^{+5.5}_{-5.4}$& &$0.441\pm0.003$&$599.0^{+5.5}_{-5.3}$\\
    0.3 &$0.459\pm0.003$&$660.1^{+6.2}_{-6.1}$& & $0.465\pm0.003$&$664.5^{+6.3}_{-6.2}$& &$0.462\pm0.003$&$661.7^{+6.2}_{-6.1}$\\
    0.0 &$0.474\pm0.003$&$717.1^{+6.8}_{-6.9}$& &$0.476\pm0.003$&$716.2\pm6.9$& &$0.477\pm0.003$&$731.7\pm7.1$\\
    \hline 
\end{tabular}
\label{volumefits}
\end{table*}

\begin{figure}
\includegraphics[trim={0.5cm 0.5cm 0.4cm 0.4cm}, clip, width=\linewidth, keepaspectratio]{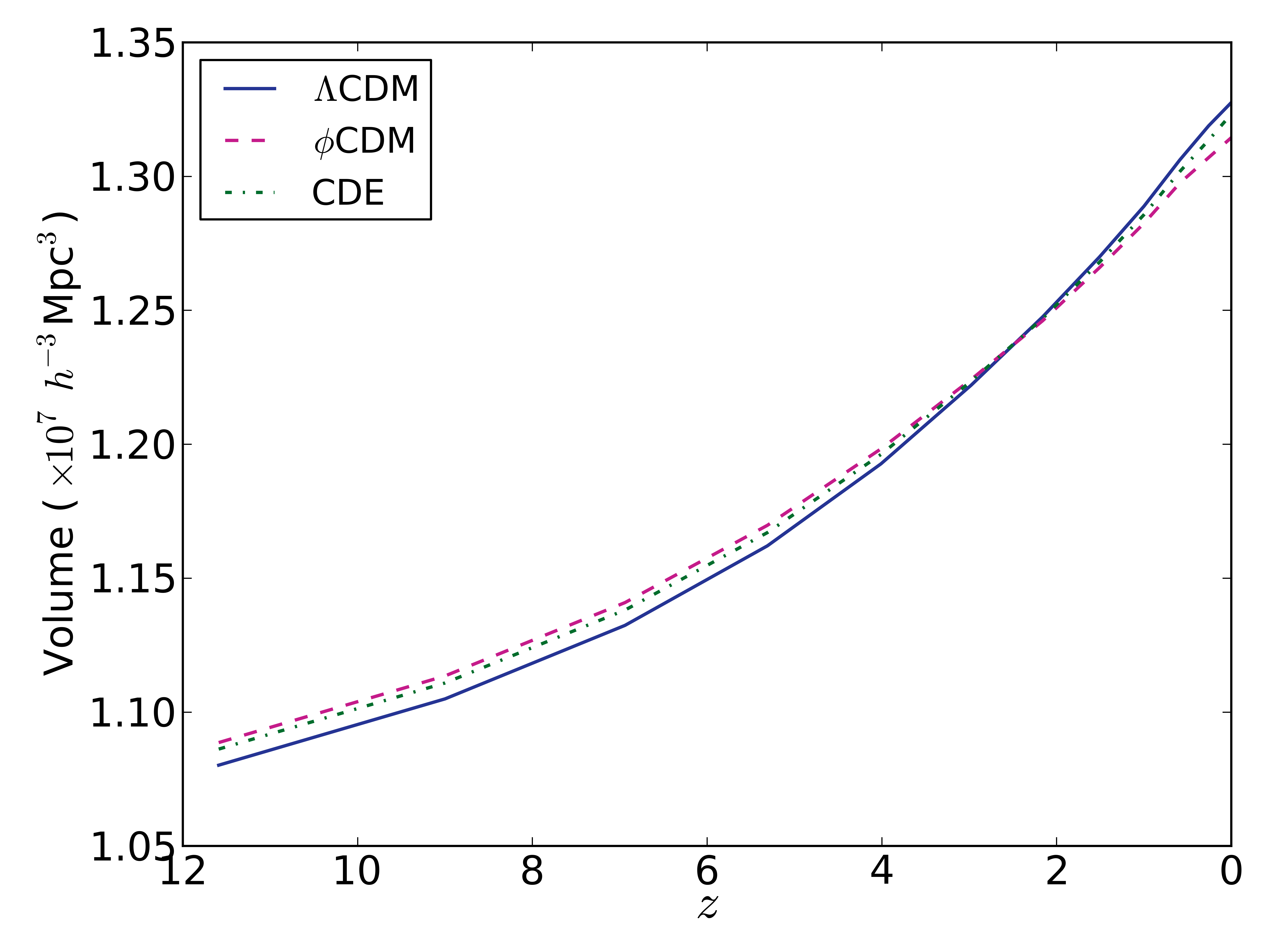}
\caption{The evolution of the total void volume in each model.} 
\label{totalvol} 
\end{figure}

Figure~\ref{totalvol} shows the evolution of the total volume occupied in voids in each cosmological model. We do not show uncertainties arising from cosmic variance, as this would require multiple simulations. Although the discrepancies may not be statistically significant for our sample size, they are still noteworthy due to our effort to minimise the effects of cosmic variance, so that discrepancies may be attributed to differences in cosmology. We find that the $\phi$CDM model contains the greatest total volume within voids at early times, while $\Lambda$CDM contains the least. At $z\approx3$, the curves intersect and $\Lambda$CDM experiences faster growth in total void volume than the other two cosmologies. At late times, $\phi$CDM contains the lowest total volume within voids. 

\subsection{Shapes}
Following \cite{2017MNRAS.468.3381A}, we fit the voids with ellipsoids and calculated their ellipticity and prolateness. To determine the axis lengths of the best-fitting ellipsoids, we calculated the moment tensor for each void using the sheet-like cells in the boundary layer:  
\begin{equation}
M_{ab} = \sum_{i} (x_{i}^{a} - X^{a})(x_{i}^{b} - X^{b}), 
\end{equation}
where $i$ represents the sheet-like cells defining the boundary layer of each void, $x^{a}_{i}$ and $x^{b}_{i}$ are the $a$-coordinate and the $b$-coordinate of the $i^{th}$ cell respectively (where $a$ and $b$ denote $x$, $y$ or $z$), and $X^{a}$ and $X^{b}$ are the $a$-coordinate and $b$-coordinate of the void barycentre respectively. The void barycentres were calculated by taking the unweighted average of the boundary cell positions. The eigenvalues of the moment tensor relate simply to the axis lengths $a$, $b$ and $c$ of the best-fitting ellipsoid: 
\begin{equation}
e_{1} = \frac{a^{2}}{3}, ~e_{2} = \frac{b^{2}}{3}, ~e^{3} = \frac{c^{2}}{3}, 
\end{equation}
where $c$ is the longest axis and $a$ is the shortest axis.

We define ellipticity of the best-fitting ellipsoid to be 
\begin{equation}
e = \frac{1}{4}\frac{c^2-a^2}{a^2 + b^2 + c^2}. 
\end{equation}
The consequence of our definition is that spheres have an ellipticity of $0$, and the value of $e$ increases as the deviation from a sphere increases. Prolateness is defined as 
\begin{equation}
p = \frac{1}{4}\frac{(b^2-a^2)+(b^2-c^2)}{a^2+b^2+c^2}, 
\end{equation}
and characterises the elongation of the axes relative to each other. A negative value of $p$ indicates that one axis is elongated relative to the other two (prolate), while a positive value of $p$ indicates that two axes are elongated relative to the third (oblate). 

\begin{figure*}
\includegraphics[width=\textwidth]{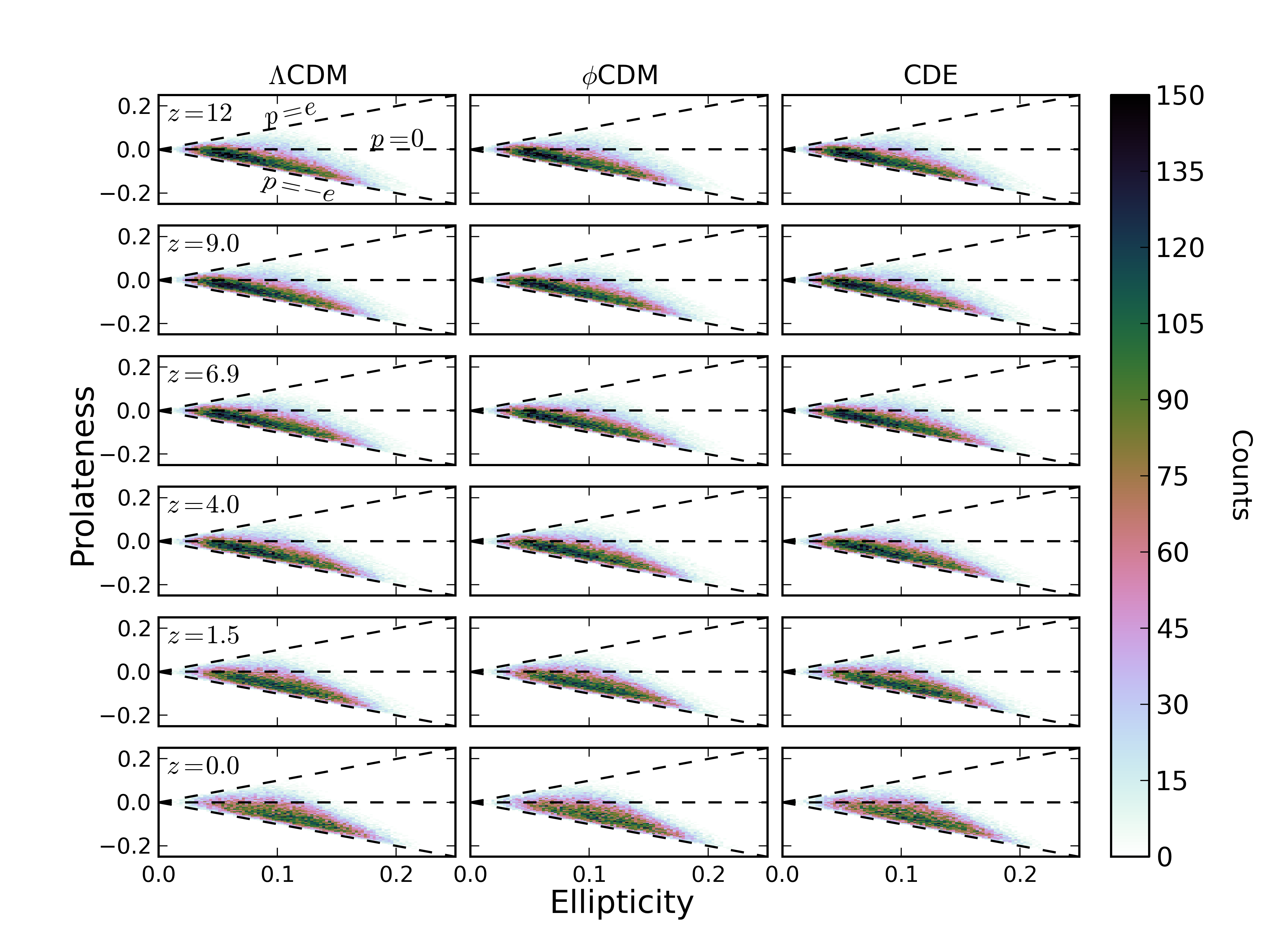} 
\caption{The distribution of ellipticities versus prolateness for the voids in a number of simulation snapshots. We show the distributions for each cosmology for the redshifts $z=0.0$, $1.5$, $3.0$, $5.3$, $6.9$, $9.0$ and $12$ (we display the $z=0.0$ distributions from \citet{2017MNRAS.468.3381A} for reference). The horizontal black dashed line denotes a prolateness of zero. The $p=\pm e$ lines represent the limiting cases where the two shortest axes are equal in length ($p=+e$), and the two longest axes are equal in length ($p=-e$).}
\label{evsp} 
\end{figure*}

Figure~\ref{evsp} shows the distribution of ellipticities and prolateness for voids in each model for selected redshifts. As the smallest voids do not have sufficiently resolved shapes, we included in our subsequent analyses only voids consisting of more than 10 cells. The majority of voids have $p<0$, and $e\approx0.1$, for $z=12-1.5$ and for all models. However, each model starts out at $z=12$ with more voids of (relatively) high prolateness or oblateness (i.e. magnitude of $p\approx0.1-0.2)$. As the models evolve, the void populations tend towards more average values of $p$ and $e$ ($p\approx-0.1-0$ and $e\approx0.05-0.15$). 

This trend can also be seen in how the median $e$ and $p$ values change with redshift. At $z=12$ the medians in $\Lambda$CDM occur at ($e$, $p$) $=$ ($0.09\pm0.03$,$-0.04^{+0.03}_{-0.04}$)\footnote{The upper and lower limits reported here are the 25$^{\mathrm{th}}$ and 75$^{\mathrm{th}}$ percentiles}. The median $e$ increases slowly with redshift, while the median $p$ decreases slowly with redshift. At $z=1.5$, the median $e$ and $p$ occur at ($e$, $p$) $=$ ($0.10\pm0.03$,$-0.05^{+0.03}_{-0.04}$), before settling at ($e$, $p$) $=$ ($0.11\pm0.03$, $-0.05\pm0.04$) when $z=0$. 

The alternative models show very similar trends overall; differences in the median values and scatter are only slight. The only differences are that the median prolateness of voids in $\phi$CDM is slightly lower than in $\Lambda$CDM, and at $z=1.5$, the scatter in the ellipticity values in CDE is slightly narrower than in the other models. The rate of change in the distributions are also extremely similar among the models. 

\begin{figure*}
\includegraphics[width=\textwidth]{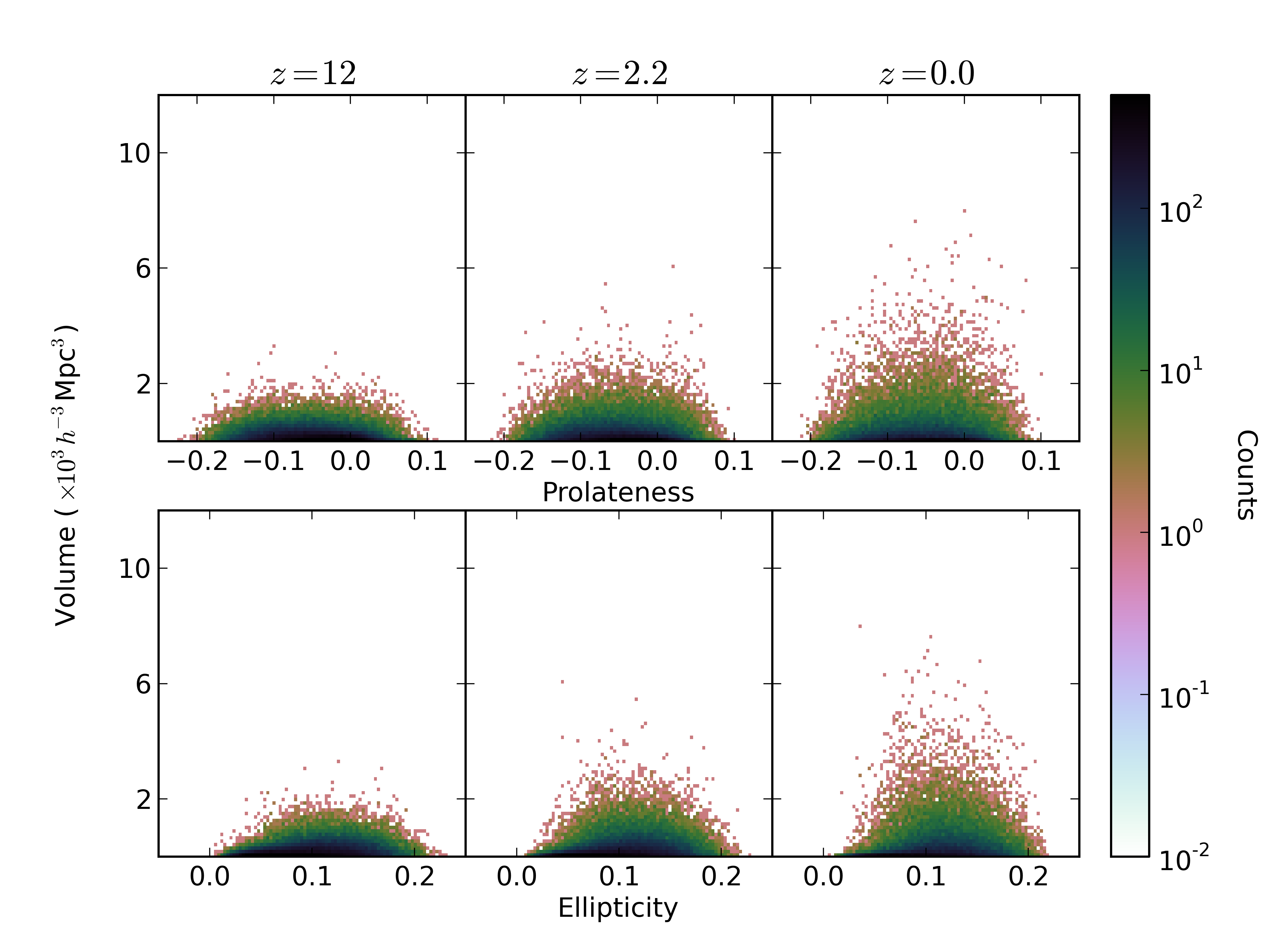}
\caption{Plots showing how ellipticity and prolateness relate to void volume in the $\Lambda$CDM model at $z=12$, $2.2$ and $0.0$. The top row shows prolateness against void volume while the bottom row shows ellipticity against void volume.} 
\label{pvsvol}
\end{figure*}

Plots of the variation in prolateness and ellipticity with volume in Figure~\ref{pvsvol} show that the oblate voids, and the voids with the highest ellipticity and lowest ellipticity in $\Lambda$CDM decline in number with decreasing redshift. Although this could be affected by small number statistics, the figure also shows that larger voids have a noticeably smaller spread in both prolateness and ellipticity than smaller voids. There is a general trend of decreasing scatter in these shape parameters with increasing volume. This is also reflected in the other two cosmological models. Additionally, there are occasional large voids tending away from the average, with positive $p$ instead of negative. However, the large void population is too small to draw any conclusion about the existence of a significant subset of large, oblate voids. 

\subsection{Average Densities}
For each model, we fit a PDF to the distribution of average void density at different redshifts. The average density for each void was calculated by taking the mean of the cell densities comprising the void, excluding the densities of the sheet-like boundary cells. For each snapshot, the average void density formed a distribution which was best fit with a skewed Gaussian,  
\begin{equation}
p(t) = \frac{1}{\sqrt{2\pi}}e^{-t^2/2}\left[1+\mathrm{erf}\left(\frac{\alpha t}{\sqrt{2}}\right)\right],
\end{equation}
where $t=(\mathrm{log_{10}}\rho -\mu)/\sigma$, $\alpha$ is the skewness parameter, $\mu$ is the mean and $\sigma$ is the standard deviation of the associated Gaussian. The best fit values for $\mu$, $\sigma$ and $\alpha$ for all available snapshots are listed in Table \ref{DensityFits}. 

\begin{table*}
\centering
\rotatebox{90}{
\begin{varwidth}{\textheight}
\begin{tabular}{c c c c c c c c c c c c}
    \hline
    &\multicolumn{3}{c}{$\Lambda$CDM} && \multicolumn{3}{c}{$\phi$CDM} && \multicolumn{3}{c}{CDE}\\
    \cline{2-4}
    \cline{6-8}
    \cline{10-12} 
  	$z$ & $\mu$ & $\sigma$ & $\alpha$ && $\mu$ & $\sigma$&$\alpha$ & &$\mu$ & $\sigma$ & $\alpha$ \\
    \hline
    \hline
 	12 & $-1.4705\pm0.0003$ & $0.0635\pm0.0003$ &$-1.9580^{+0.0242}_{-0.0003}$ &&$-1.4814\pm0.0004$&$0.0653\pm0.0003$&$-1.7328^{+0.0231}_{-0.0003}$&& $-1.4770\pm0.0003$& $0.0647\pm0.0003$& $-1.8112^{+0.0233}_{-0.0003}$\\
  	9.0 & $-1.4910\pm0.0005$& $0.0680\pm0.0004$& $-1.4970^{+0.0229}_{-0.0004}$&&$ -1.5053\pm0.0005$&$0.0698\pm0.0004$&$-1.3010^{+0.0229}_{-0.0004}$&& $-1.4992\pm0.0005$& $0.0691\pm0.0004$&$-1.3750^{+0.0223}_{-0.0004}$\\
    6.9 & $-1.5179\pm0.0007$& $0.0732\pm0.0005$& $-1.1162^{+0.0245}_{-0.0005}$&&$-1.5373\pm0.0010$&$0.0746\pm0.0006$&$-0.9155^{+0.0272}_{-0.0006}$&&$-1.5282\pm0.0008$&$0.0743\pm0.0005$&$-1.0110^{+0.0255}_{-0.0005}$\\
    5.3 &$-1.5568^{+0.0015}_{-0.0016}$& $0.0775^{+0.0007}_{-0.0008}$& $-0.7072^{+0.0360}_{-0.0008}$&&$-1.6571^{+0.0015}_{-0.0014}$&$0.0866\pm^{+0.0008}_{-0.0009}$&$0.8858^{+0.0335}_{-0.0009}$ &&$-1.5751^{+0.0027}_{-0.0035}$&$0.0765^{+0.0011}_{-0.0012}$&$-0.5032^{+0.0682}_{-0.0012}$\\
    4.0 &$-1.6879\pm0.0012$& $0.0993^{+0.0007}_{-0.0008}$& $1.0883^{+0.0291}_{-0.0008}$&&$-1.7214\pm0.0010$&$0.1107\pm0.0007$& $1.2963^{+0.0275}_{-0.0007}$&&$-1.7054\pm0.0011$&$0.1054\pm0.0007$&$1.2073^{+0.0281}_{-0.0007}$\\
    3.0 &$-1.7554\pm0.0010$& $0.1248\pm0.0008$&$1.4656^{+0.0281}_{-0.0008}$&&$-1.7905\pm0.0010$&$0.1359\pm0.0008$&$1.6167^{+0.0284}_{-0.0008}$&&$-1.7731\pm0.0010$ &$0.1308\pm0.0008$&$1.5545^{+0.0280}_{-0.0008}$\\
    2.2 &$-1.8296\pm0.0010$& $0.1522^{+0.0008}_{-0.0009}$& $1.8143^{+0.0306}_{-0.0009}$&&$-1.8674\pm0.0010$&$0.1639\pm0.0009$&$1.9251^{+0.0311}_{-0.0009}$&&$-1.8472\pm0.0010$ &$0.1574\pm0.0009$&$1.8596^{+0.0306}_{-0.0009}$\\ 
    1.5 &$-1.9093\pm0.0011$&$0.1800\pm0.0010$& $2.0624^{+0.0328}_{-0.0010}$&&$-1.9490\pm0.0011$&$0.1917\pm0.0010$&$2.1827^{+0.0337}_{-0.0010}$ &&$-1.9275\pm0.0011$&$0.1852\pm0.0010$&$2.1147^{+0.0335}_{-0.0010}$\\
    \hline
    1.0 &$-1.991\pm0.001$&$0.205\pm0.001$&$2.256^{+0.036}_{-0.001}$&&$-2.033\pm0.001$&$0.217\pm0.001$&$2.368^{+0.037}_{-0.001}$&&$-2.010\pm0.001$&$0.210\pm0.001$&$2.295^{+0.036}_{-0.001}$\\
    0.6 &$-2.071\pm0.001$&$0.226\pm0.001$&$2.415^{+0.039}_{-0.001}$&&$-2.113\pm0.001$&$0.237\pm0.001$&$2.493^{+0.041}_{-0.001}$&&$-2.092\pm0.001$&$0.232\pm0.001$&$2.474^{+0.041}_{-0.001}$\\	0.3&$-2.147\pm0.001$&$0.248\pm0.001$&$2.580^{+0.043}_{-0.001}$&&$-2.177\pm0.001$&$0.255\pm0.001$&$2.644^{+0.043}_{-0.001}$&&$-2.168\pm0.001$&$0.252\pm0.001$&$2.601^{+0.043}_{-0.001}$\\     
    0.0 &$-2.208\pm0.001$&$0.260\pm0.001$&$2.600^{+0.044}_{-0.001}$&&$-2.251\pm0.001$&$0.272\pm0.001$&$2.715^{+0.046}_{-0.001}$&&$-2.246\pm0.001$&$0.270\pm0.001$&$2.700^{+0.047}_{-0.001}$\\
    \hline 
\end{tabular}
\caption{The best fit values (with $1\sigma$ uncertainties) for the parameters $\mu$, $\sigma$ and $\alpha$. The values for $z=0.0-1.0$ are also included for the sake of comparison.} 
\label{DensityFits} 
\end{varwidth}}
\end{table*}

\begin{figure*}
\includegraphics[width=\textwidth]{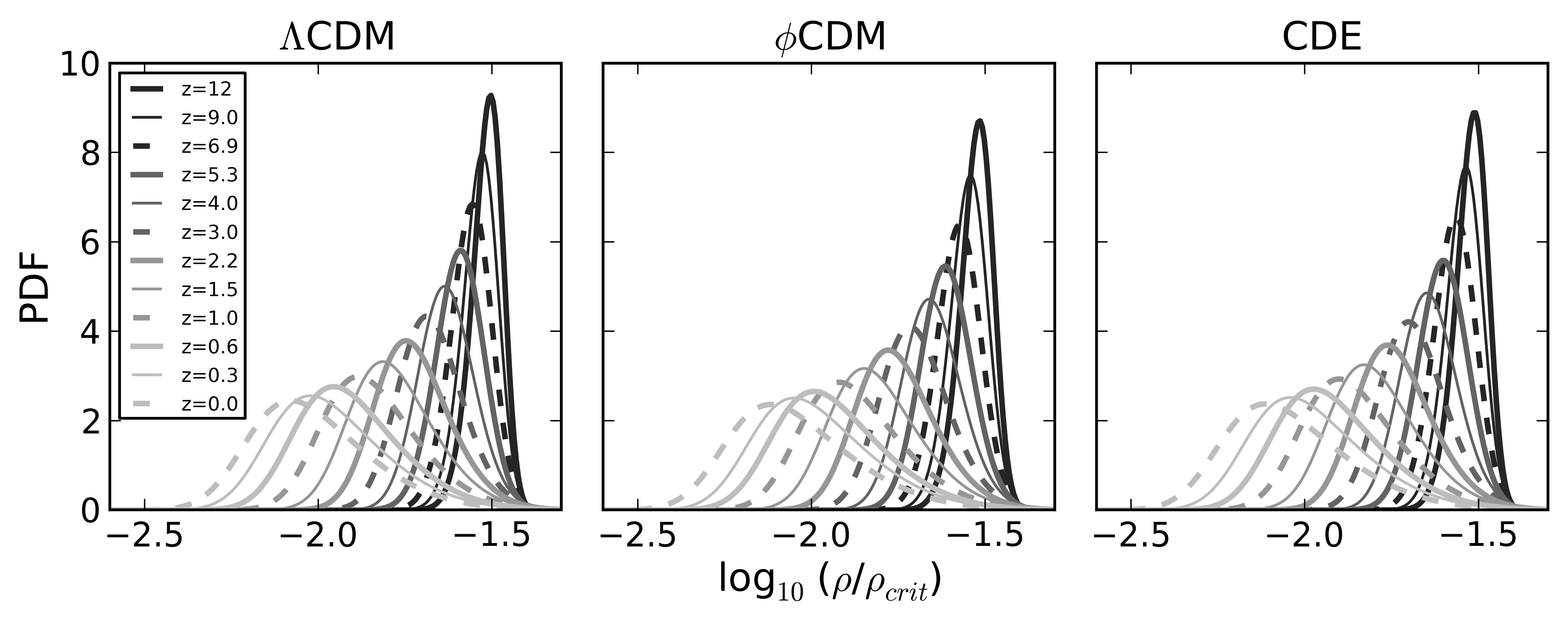}
\caption{Evolution of the average density PDFs from $z=12$ to $z=0.0$.} 
\label{densPDF}
\end{figure*}

The probability density functions showing the average void density distribution from $z=12-0$ for each cosmological model are displayed in Fig~\ref{densPDF}. The peak of the PDFs decrease with time, showing that voids become more underdense with time. The spread of the PDFs also increases with decreasing redshift, suggesting that the range of densities starts off quite narrow at early times and increases as voids evolve (expand, evacuate and merge). The evolution of the density PDFs in all three models is very similar. 

\begin{figure*}
\includegraphics[width=\textwidth]{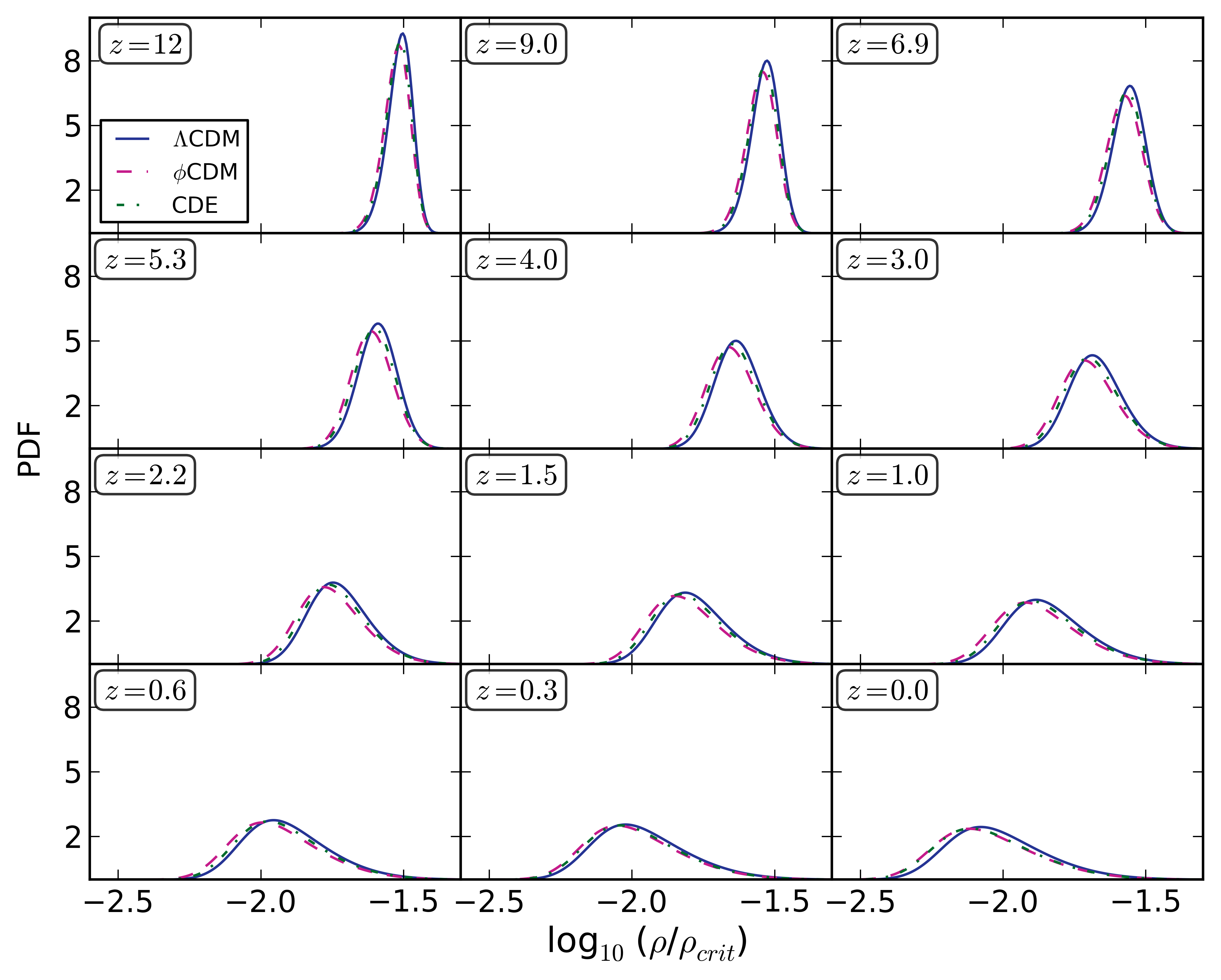}
\caption{Comparison of the density PDFs between the three models at multiple redshifts.} 
\label{densPDF2}
\end{figure*}

Figure~\ref{densPDF2} shows the comparison between the density PDFs at specific redshifts. The density PDFs at $z=0-1$, which were presented in \cite{2017MNRAS.468.3381A}, are also displayed in this figure for completeness. At $z=12$, the three models show the most similarity in their density distributions, but remain distinct. The $\Lambda$CDM density PDF peaks at a slightly higher average void density than the other two models. As the models evolve, the $\Lambda$CDM PDF becomes more distinct from, and continues to peak at a higher average density than, the other two PDFs. At $z=5.3$, the $\phi$CDM PDF starts to become more distinct from the CDE PDF, shifting further to the left and showing a wider spread in densities. This indicates that the $\phi$CDM density PDF starts spreading out to include lower average void densities a sooner than the PDFs of CDE and $\Lambda$CDM. From $z=4.0$ to $z=1.0$, the three models are quite distinct from each other, with $\phi$CDM peaked at the lowest void density and with the widest spread, and $\Lambda$CDM peaked at the highest and with the narrowest spread. The $\Lambda$CDM and CDE PDFs finally attain a similar spread to the $\phi$CDM PDF by $z=0.6$, and as observed in \cite{2017MNRAS.468.3381A}, the CDE PDF approaches the $\phi$CDM PDF until they converge at $z=0.0$. 

\begin{figure*}
\includegraphics[trim={3.0cm 0.5cm 3.5cm 1.3cm}, clip, width=\textwidth]{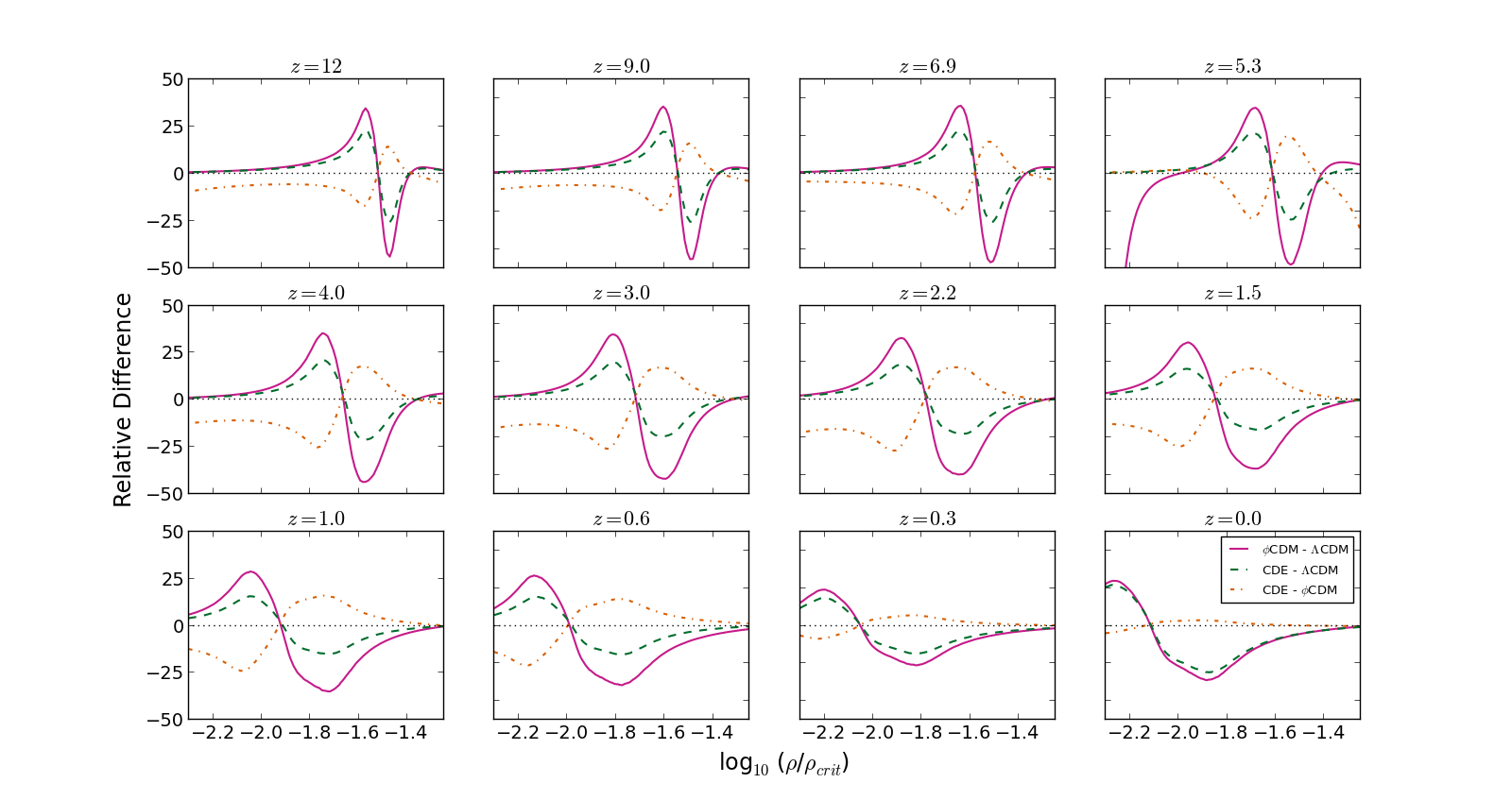}
\caption{The relative differences in the density PDFs between each pair of the three models, $\phi$CDM and $\Lambda$CDM (solid magenta line), CDE and $\Lambda$CDM (dashed green line), and CDE and $\phi$CDM (dotdashed orange line), from $z=12-0$ (later redshifts included for reference). The relative differences were calculated using the median ratios and are expressed in units of half the range between the 16$^{\mathrm{th}}$ and 84$^{\mathrm{th}}$ percentiles. A relative difference of zero is indicated by the dotted black line.} 
\label{DensityBootstraps}
\end{figure*}

To compare the density PDFs between the models, we performed a bootstrapping analysis on the distribution of parameter values that were sampled during the MCMC fitting process. As with the volume PDF comparisons, we took a subsample of 5000 values for each of the three parameters, $\mu$, $\sigma$ and $\alpha$, and calculated the ratios between the corresponding volume PDFs they defined, thus producing a distribution of ratios between each pair of models. From these ratios, we calculated the relative difference between each pair of cosmologies, which we define as $(\mathrm{Model 1}/\mathrm{Model 2} -1)/\sigma_{0}$, where $\mathrm{Model 1}/\mathrm{Model 2}$ is the median ratio, and $\sigma_{0}$ is equivalent to half the range between the 16$^{\mathrm{th}}$ and 84$^{\mathrm{th}}$ percentile ratios. The relative differences are presented in Figure~\ref{DensityBootstraps}, highlighting the statistically significant differences between the PDF fit parameters and shapes. 

We note that representing the relative differences in terms of a $\sigma_{0}$ defined in this way does not account for the asymmetry in the upper and lower uncertainties associated with the relative difference (or ratio) values. However, since the differences between the density PDFs are very significant, these asymmetries do not have much effect on the accuracy of our representation, while clearly showing the extent of the deviations between the models. \footnote{We did not use relative difference to represent deviations between volume PDFs in Figure \ref{volPDFs}. This is because the asymmetries about the median in the 16$^{\mathrm{th}}$ to 84$^{\mathrm{th}}$ percentile range were critical for the accurate representation of the differences and their statistical significance.} 


It is clear from the figure that across all redshifts (except at $z=0$) the alternative models' average void density distributions are inconsistent with that of the standard model, with much more than a $1 \sigma_{0}$ difference between their distributions. The only crossings with a relative difference of zero occur where the PDFs intersect each other. We note that at $z=0$, the relative differences between the CDE and $\phi$CDM PDF appear to be greater than $1 \sigma_{0}$ for some densities in Figure~\ref{DensityBootstraps}, but this is only an artefact of our choice to define $\sigma_{0}$ as half the range between the 16$^{\mathrm{th}}$ and 84$^{\mathrm{th}}$ percentiles, not accounting for asymmetry in the ratio distribution and thus slightly underestimating the size of the upper or lower uncertainty in the median ratio at certain densities. The differences between the CDE and $\phi$CDM PDFs are not statistically significant, as reported in \cite{2017MNRAS.468.3381A}. Although the relative differences reach $> 25 \sigma_{0}$ in significance at most redshifts, the PDF shapes do not wildly differ from each other, as seen in Figure~\ref{densPDF2}. We also note that the relative differences between models at $z=5.3$ exhibit a different shape than those at other redshifts, because each density PDF undergoes a transition from negative to positive skewness at approximately this redshift, but not at exactly the same time.  

\begin{figure*}
\includegraphics[width=\textwidth]{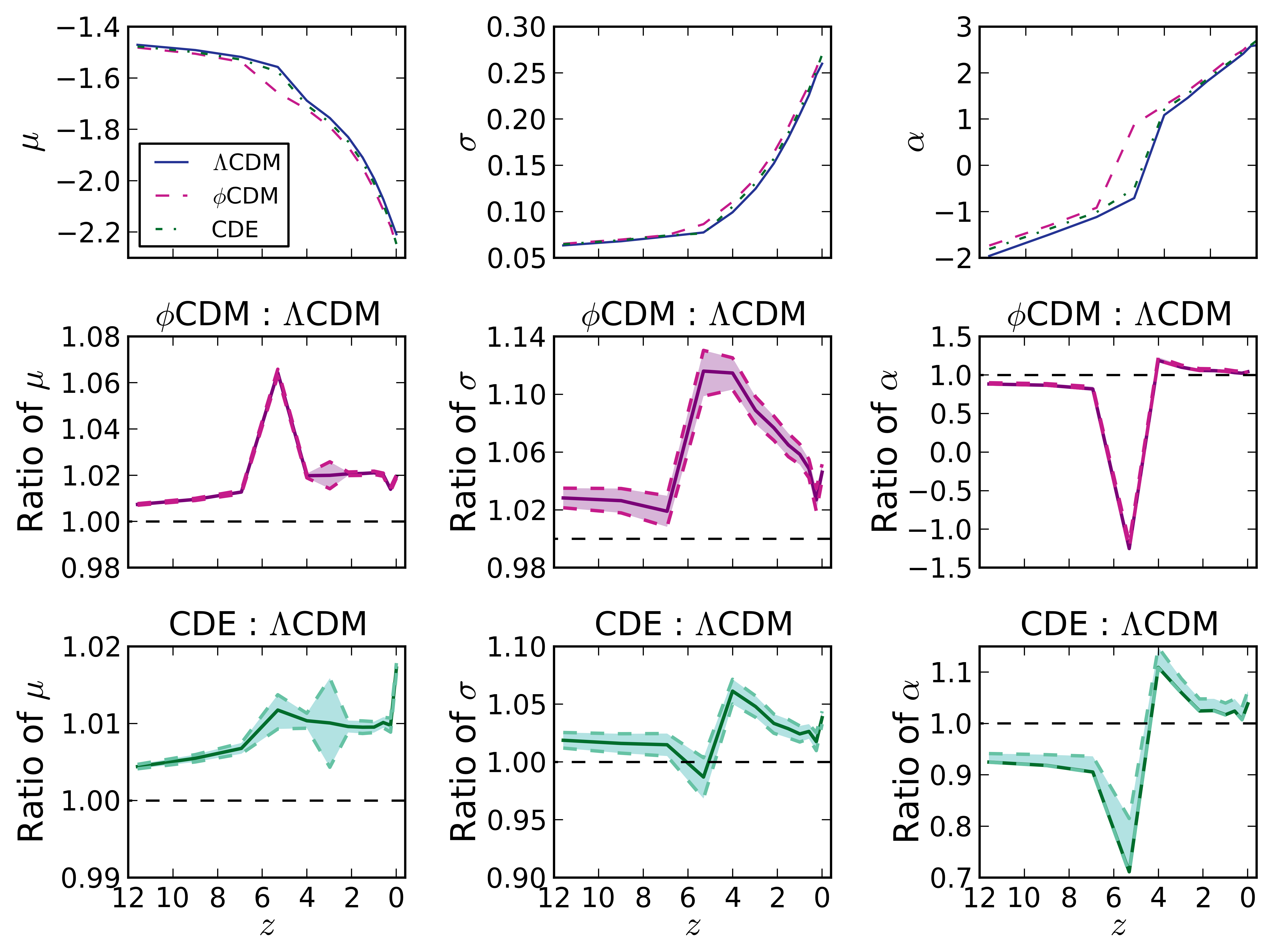}
\caption{Evolution of density fit parameters and their ratios with $1 \sigma$ uncertainties, from $z=12-0$. The evolution of the best fit values in each of the models is displayed in the first row. The ratio between the $\Lambda$CDM best fit values and those of the $\phi$CDM and the CDE models are displayed in the second and third row, respectively. A ratio of unity is represented by the black dashed line.}
\label{densparams}
\end{figure*}

In Figure~\ref{densparams}, we show the evolution of the best fit density PDF parameters, $\mu$, $\sigma$ and $\alpha$, across all models and all available redshifts (for completeness, we also display the best fit values from $z=1-0$). As with the volume parameter ratios, the uncertainties were calculated by adding the parameter uncertainties in quadrature, accounting for asymmetry in the upper and lower uncertainties. All three models show a decrease in $\mu$, and increases in $\sigma$ and $\alpha$, with time, indicating greater underdensities and larger spreads in densities with time. The CDE and $\Lambda$CDM best fit values show the greatest similarity, while $\phi$CDM shows the greatest deviation from the other two models, particularly in $\mu$ and $\alpha$ and from $z=7$ to $z=4$. 

The $\phi$CDM PDF changes shape earlier than the other two PDFs. Specifically, the skewness changes from negative to positive ($\alpha$ changes sign) earlier, and the PDF experiences an earlier and faster decline in the peak density from $z\approx7-5$ before slowing down again. The $\Lambda$CDM and CDE PDFs experience that faster decline from $z\approx5-4$. As can be seen in the second and third rows of Figure~\ref{densparams}, the $\phi$CDM and CDE parameters are largely distinct from those of $\Lambda$CDM. 

\section{Discussion}\label{Discussion}
A key element in the interpretations of our results is the fact that our simulations begin with matching density perturbation phases, allowing for initial over- and under-densities to seed in the same location independent of cosmology. This means that differences among the simulations are due to the differences in expansion history and the growth of the density perturbation amplitudes, hence underlying cosmology, rather than cosmic variance. The differences we focus on in this section are at least $1\sigma$ deviations. However, due to the phase matching and enforcing the same $\sigma_{8}$ at $z=0$, any systematic differences that are within a $1\sigma$ consistency are still informative, especially if they do not decrease with redshift. 
\subsection{Population Size}
The fact that the void population size is similar across the models for all redshifts shows that the primary determinant of void formation and growth is gravitational interactions (present in all models). However, given that these simulations start with the same initial density perturbation phases (which allow the same voids to appear in each simulation at the beginning), the small discrepancies we observe in the population sizes are notable, and likely to be directly linked to differences in dark sector physics. 

The decline in population size with time for all three models is due to the merger of smaller voids (initially mildly non-linear) into larger voids. The general decline in the $\phi$CDM and CDE void population sizes compared to the standard model could be due to differences in void merger rates between the models. In particular, the alternative models appear to have a greater void merger rate than the standard model, resulting in these models containing fewer voids than the standard model at late times. This increased void merger rate may be due to a higher particle evacuation rate from voids in the alternative models associated with the dynamical scalar field \citep[proposed in][]{2017MNRAS.468.3381A}. The wall of particles between two nearby voids would evacuate more quickly, enabling the two voids to merge earlier than they would under the standard model. A greater void evacuation rate could also reasonably lead to faster void formation and thus more voids at higher redshifts compared to $\Lambda$CDM, as seen in Figure~\ref{voidcounts}. 

At $z\approx0.3-0.6$, we observe a sudden increase in the population ratios. This could be the result of a change in the relative merger rates between the alternative models and the standard model somewhere in this redshift range, either due to changes in the $\Lambda$CDM merger rate or changes associated with the scalar field. More in-depth investigation into merger rates is needed to pinpoint the exact cause of the increase. 

The drag force on baryons due to dark sector coupling is the likely reason for the less pronounced deviation between the CDE and $\Lambda$CDM void populations, as the additional drag force would slow down void evacuation rates and thus merger rates compared to $\phi$CDM. However, the sudden decline in the CDE void population compared to $\phi$CDM at very late times, suggests an additional effect is at play. This decline occurs when the average density distribution of CDE voids approaches that of $\phi$CDM voids, and when its large void population ($V > 1000$ $h^{-3}$Mpc$^{3}$) becomes significantly greater than the other two models \citep[see][]{2017MNRAS.468.3381A}. One likely explanation is a sudden increase at very late times in the evacuation rate of the CDE voids compared to $\phi$CDM, which would reduce the average void densities and increase the merger rate in the CDE model, explaining both the greater number of large voids and the drop in void population. However, it is unclear how coupling might increase the evacuation rate at very late times. Further study into void merger rates and how they vary with redshift and model is required to explain these changes in the CDE model (Adermann et al. in prep). 

\subsection{Volume Distributions}
The volume distributions seen in Figure~\ref{volPDFs} are very similar in shape to each other and to the PDFs presented in \cite{2017MNRAS.468.3381A}. We observe that the PDFs turn over at larger volumes as redshift decreases, consistent with a progressive increase in steepness of the power law $\alpha$ and the characteristic volume $V_{0}$. The curves show that at higher redshifts for all three models, the probability of smaller voids in any given volume is higher than at lower redshifts, while the probability of larger voids is lower than at lower redshifts. These trends are consistent with what we expect of voids as they merge and expand, which would increase the number of large voids with time while decreasing the number of small voids. However, the trends also show that the rate at which the mid-range voids are replenished is not as great as the rate at which large voids are created. The smallest of voids ($2$ $h^{-3}$Mpc$^{3}$ $\leq V \leq 4$ $h^{-3}$Mpc$^{3}$) tend to become more numerous with decreasing redshift, suggesting that the rate at which they are growing/merging into mid-range voids is less than the rate at which they are forming. Their existence at low redshift shows that small voids appear throughout the evolution of the universe, regardless of model. 

The comparisons between the volume PDF parameter values in Figure~\ref{volPDFimage} and Table~\ref{volumefits} show that $\alpha$ and $V_{0}$ are almost all consistent with one another between models, and follow the same trend with time. It is clear from the rate of increase in $V_{0}$ that the birth rate of larger voids increases with time, meaning that the total void expansion and merger rate increases with time for all three models. The question of which of the two methods for growing voids largely determines the rate of large void formation will be investigated in later publications, although since mergers are a faster method of growing voids than void wall expansion, we believe that it is dominated by the merger rate. However, the discrepancy ($>1\sigma$) at $z=0$ between the CDE value of $V_{0}$ and the $\Lambda$CDM value reveals that large voids have a faster growth rate at very late times under a CDE cosmology. This is consistent with the population size results, from which we concluded that CDE must have a greater merger rate at very late times in order to have the smallest population size by $z=0$. Thus, we propose that the increased growth rate of large voids in CDE at late times is the result of an increased merger rate, which in turn is due to an increased particle evacuation rate from voids at late times. This could in fact be the primary cause of the deviation we observed in the CDE volume PDF at $z=0$ in \cite{2017MNRAS.468.3381A}, where the CDE model predicted a greater number of voids at high volumes compared to the other two models. 

From the lack of statistically significant deviation between the alternative and standard model volume PDFs for large voids ($V > 1000$ $h^{-3}$Mpc$^{3}$), across $z=1.5-12$, we can conclude that the growth of voids and their eventual sizes are primarily governed by gravitational effects. However, although not statistically significant, the differences in the median void volumes between the alternative and standard models suggests some small dependence of void volume/growth on cosmology. For example, CDE contains the most large voids at the highest redshifts, until $z\approx5.3$, when $\phi$CDM then contains the most (see Figure~\ref{volPDFs}). This can also be explained by a greater void merger rate in the alternative models compared to the standard model, as void merger rate is positively correlated with the formation rate of large voids. However, this does not explain why CDE starts at $z=12$ with more large voids than $\phi$CDM. It is possible that the coupled dark sector leads to greater numbers of large voids at early times, perhaps due to a higher merger rate and/or a higher void wall expansion rate at early times, but the drag force associated with coupling slows down their growth rate so that at later times, $\phi$CDM contains more large voids. 

The fact that the $\Lambda$CDM model contains the largest number of mid-sized voids (those with $100 \lesssim V \lesssim 1000$ $h^{-3}$Mpc$^{3}$) from $z=1.5-12$ can also be explained by the lower merger rate in $\Lambda$CDM. A lower merger rate would slow down the creation of large voids more than the creation of mid-sized voids, which can also be formed through void wall expansion. Interestingly, there is no clear trend for the smallest voids ($V \lesssim 100$ $h^{-3}$Mpc$^{3}$), with different models containing the greatest number of these voids at different redshifts. This suggests that the growth and depletion of the smallest voids occurs at different rates in each model at different times. 

Although the differences we see in the volume distributions suggest differences in void growth rates, stemming from differences in underlying cosmology, we require confirmation of this explanation with further studies of how void merger and void expansion rates differ among the three models, which will be presented in future publications (Adermann et al., in prep.). 

In \cite{2017MNRAS.468.3381A}, we suggested that there could be greater discrepancies in the volume PDF between $\phi$CDM and $\Lambda$CDM at higher redshifts (because the trend from $z=1-0$ showed a decline in the discrepancy towards $z=0$), however we observe that there is no such continuing trend into the higher redshifts. Instead, the ratio between the $\phi$CDM and $\Lambda$CDM volume PDFs rises and falls with time. Without further data, we cannot determine the cause of this. Furthermore, these fluctuations are well within the $1\sigma$ uncertainty range, and may well be statistical (for example, from the lack of large-scale power in the simulations) rather than meaningful fluctuations.  

Finally, we note that $\phi$CDM starts out with the greatest total volume and $\Lambda$CDM the least total volume at early times, and then this crosses over at $z\approx2-3$ and the opposite order appears (see Figure~\ref{totalvol}). At early times, the $\phi$CDM model has the highest void population (see Section~\ref{Popsize}), and unsurprisingly, the greatest total volume contained in voids. Despite having fewer large voids than the other two models, $\Lambda$CDM has the highest void population and the greatest total void volume at late times, which is likely due to having more mid-sized voids than the alternative models. At $z=0$, CDE has the smallest void population and yet it has a greater total void volume than $\phi$CDM. This is consistent with the result from \cite{2017MNRAS.468.3381A}, showing that CDE has more large voids than the other two models, which very likely were the result of increased merger rates, and potentially even an increased void wall expansion rate. These large voids would contribute to the total void volume without adding to the population. 

\subsection{Shape Distributions}
From $z=12-1.5$, void growth/merger dynamics result in only small and slow changes to the void ellipticity and prolateness distributions. Nevertheless, some of these changes may offer insight into the evolution of voids. Firstly, we observed that for all three models, there is a general decrease in the numbers of high ellipticity and low ellipticity voids with time\footnote{We define high ellipticity voids as those with $e\approx0.2$, which is high compared to the vast majority of the population (but not compared to the full range of ellipticities). Low ellipticity refers to $e\approx0$.} (Figure~\ref{evsp}). Secondly, larger voids tend to stay approximately centred around the median values while the scatter in their prolatenesses and ellipticities reduces. Despite the small number statistics and lack of large-scale modes, the existence of these trends in all three models suggests that it is not merely a statistical fluke. However, we acknowledge that since our data does not confirm these trends in simulations with different initial density perturbations, it is possible that these results are not robust under cosmic variance. 

Whilst many small voids start out with high ellipticities, when they merge they are much more likely to produce a shape with less extreme values of $e$ and $p$. This trend is a consequence of the greater number of ways to combine voids into less elliptical shapes, as high ellipticity voids require more specific and less likely alignments of voids before merging. The tendency towards shapes with ($e$, $p$) $\approx$ ($0.1$,$-0.05$), even at the highest redshifts, is not as straightforward. Our results suggest that voids tend to evacuate faster along one direction (producing prolate voids) rather than two or three directions (which would produce oblate or non-elliptical voids), which is consistent with `Zel'dovich pancake' collapse \citep[see][]{2014MNRAS.441.2923C}. Voids do not tend to evacuate at the same rate in all three orthogonal directions in our simulations, which is perhaps not surprising because the cosmic web is not anisotropic along all directions for an observer within a void. However, this does not fully explain why the preferred prolateness is around $-0.05$, although it could be due to mergers pushing voids towards lower magnitudes of $p$. It is beyond the scope of our paper to elicit how mergers and void growth affect void shapes. To answer this question, we require the shape distributions for multiple simulation volumes with differing initial density perturbations, as well as void tracking to identify the interplay between mergers and evacuation in void growth and evolution. 

The trends discussed so far do not vary with model. Thus, we conclude that the average shape distributions and the processes that produce them are generally insensitive to underlying cosmology, and are largely determined by gravity. This may not be surprising considering that void shape is defined by the boundary of the void, which is classified as sheet material, and is thus somewhat non-linear in its growth. Despite this, there do exist distinct differences in the shapes of the largest voids between the cosmological models. These are differences in the growth and evolution of individual voids, which are noteworthy because the simulations start out with the same initial density perturbation phases. Any alterations in their growth and evolution is due to differences in cosmology, and so it would be useful to isolate these differences for further investigation in void-by-void comparison studies. There may be further individual differences amongst the smaller voids that are not visible in the shape distributions due to the sheer number of similarly-sized and -shaped voids, but would be observable individually, which could offer further insight into the effect of dark sector physics on the growth and evolution of voids. 
\subsection{Average Density Distributions} 
The density evolution observed in Figure~\ref{densPDF} and Figure~\ref{densparams} is consistent with our understanding of how voids evolve under gravity. At high redshift, voids have a much narrower range of average densities. They generally become less dense due to their increasing size (from expansion and mergers) and continual evacuation of matter as they accumulate onto denser structures. Since the rate of emptying depends on the specific local structure around each void, and thus varies across the population, the spread of average densities increases over time. This gravity dominated evolution is evident in all three models. 

However, dark sector physics does leave a potentially observable imprint on the evolution of void densities, which is clear from Figures~\ref{densPDF2} and \ref{DensityBootstraps}. The three models predict distinct density PDFs at all redshifts except for $z=0$ \citep[the $z=0$ case was discussed in][]{2017MNRAS.468.3381A}. In particular, the $\phi$CDM density distributions consistently peak at lower densities than the other models, while $\Lambda$CDM consistently peaks at higher densities. It is clear that on average, the $\phi$CDM cosmology produces the emptiest voids and $\Lambda$CDM produces the densest, while CDE is somewhere in between. This phenomenon occurs throughout the entire evolution, from at least $z=12$ to $z=0$. Although we have not calculated density profiles (as the majority of our voids are not spherical) and determined the exact effect of the scalar field on their shapes, we expect the increased evacuation rate in our quintessence models to result in statistically significant lowered densities across most of the void, with greatest discrepancy from $\Lambda$CDM occurring at the void centre. This is consistent with the density profile results presented in \cite{2016MNRAS.455.3075P} for their coupled dark energy model at low redshift, in which greatest discrepancy from the standard model was found in the centre.  

Since the observed deviation does not shrink with time, we conclude that it is not the result of normalisation, and that the scalar field is leaving this imprint on the average density of voids. Average density is affected by both the size and evacuation rate of voids. However, the lack of significant differences in the void volume distribution among the models suggests that this is not the primary cause of lower density voids in $\phi$CDM. Instead, we propose that the scalar field affects their evacuation rate. Additionally, as dark energy in CDE and $\phi$CDM arise from the same mechanism (dynamical scalar field) and differ in the degree to which dark matter and the scalar field are coupled, we can conclude that the coupling and its associated effects cause the voids in the CDE model to be more dense than those in the $\phi$CDM model, while the scalar field forces voids to be less dense than those in the $\Lambda$CDM model. The coupling leads to a drag effect on the dark matter particles, which slows down their evacuation from voids. 

The differences between $\phi$CDM and the other two models is also apparent in Figure~\ref{densparams}. In particular, the shape of the $\phi$CDM PDF evolves more quickly than in the other two models, with the change of skewness sign and the drop in peak density occuring at earlier times, due to faster evolution and evacuation of voids. Interestingly, the shape of the parameter evolution curves for CDE and $\Lambda$CDM are remarkably similar, despite the parameter values being different. The decline in peak density happens at very similar rates, as well as the rise in the overall spread of the PDFs, and the change in skewness. The effect of coupling seems to override the effect of the scalar field when it comes to the overall parameter evolution, and thus shape evolution, bringing it much closer to a $\Lambda$CDM cosmology than a $\phi$CDM cosmology, despite the shifted density PDFs. 

Although the dynamical scalar field characterised by the Ratra-Peebles potential leaves a statistically significant and potentially observable imprint on the average void densities, the differences we see may be fairly model-dependent. Another potential could produce a different imprint (including no imprint), on void density. However, most simple scalar field models are required to reproduce the approximate expansion history of $\Lambda$CDM. Hence the scalar field density, $\Omega_{\phi}$, must exhibit late-time growth. This typically results in a faster rate of growth for $\Omega_{\phi}$ than $\Omega_{\Lambda}$, so if this is a major component of the scalar field's influence over the density of voids, then the imprint we observe could be considered a generic feature of any scalar field cosmology consistent with observations. If this is the case, the observed imprint in the average void density PDFs across a large range of redshifts would be a very promising probe of a dynamical scalar field. 

\section{Conclusions}\label{Conclusions}
We calculated and compared the void properties over the redshift range $z=1.5-12$ between the $\Lambda$CDM model of cosmology and two alternative models, $\phi$CDM and CDE. Specifically, we compared the size of the void population, the total population volume, the volume distribution, the shape distributions and the average density distributions. 

These properties were derived from three adiabatic hydrodynamical simulations (one for each cosmological model), each of which contained $512^{3}$ dark matter and baryonic particles, in a box of length $500$ $h^{-1}$Mpc. The simulations initially had the same density perturbation phases, so population differences from cosmic variance are not present. Any differences seen thus arise from differences in underlying cosmology. 

We used a Hessian-based void finder to identify voids in the cold dark matter distribution of these simulations. We examined the size of the void population and the total volume occupied by voids, their volumes, shapes and densities across cosmic time, in several cosmologies to identify how cosmology, specifically dark sector physics, affects void evolution and growth. 

Firstly, we found that the void population size and its evolution were fairly similar among the three models, which serves as evidence that gravitational effects have the greatest influence on void formation and growth. However, although not statistically significant, there exist a number of small discrepancies between the standard and alternative models, which likely arise from differences in cosmology. The alternative models contained void populations which were greater than that of $\Lambda$CDM at early times, and smaller at late times. From this, we concluded that the void merger rate within $\phi$CDM and CDE is greater than in $\Lambda$CDM. We propose that this is due to the increased evacuation rate from voids due to the scalar field, giving rise to more depleted voids in the alternative models than $\Lambda$CDM. Furthermore, we propose that the CDE cosmology experiences a sudden increase in its void evacuation rate at very late times, leading to the sudden decline in its void population, the sudden increase in the number of large voids with $V > 1000$ $h^{-3}$Mpc$^{3}$, and the sudden consistency between its average void density distribution and that of $\phi$CDM we have observed at $z=0$. 

Secondly, we found that the void volume distribution cannot distinguish between the models at the $1\sigma$ level. The presence of the scalar field or dark sector coupling does not leave a distinct imprint on the volume distribution of voids. Although not statistically significant, slight variations exist between the models. In particular, CDE and $\phi$CDM both contain more large voids (in the range $V>1000$ $h^{-3}$Mpc$^{3}$) than $\Lambda$CDM. Furthermore, the number of large voids in the CDE model relative to $\Lambda$CDM and $\phi$CDM declines with time, and is surpassed by $\phi$CDM so that by $z=1.5$, $\phi$CDM contains the highest population of large voids. We suggested that the coupling present in the CDE model has the effect of increasing the number of large voids at very early times. The very late time increase in the number of large voids in the CDE model could be the result of a sudden increase in the evacuation rate of particles from voids and hence merger rates, leading to the sudden lowering of average void density at $z=0$. We also discovered that the $\Lambda$CDM model contains the greatest number of mid-sized voids with volumes of $\sim100$ $h^{-3}$Mpc$^{3}$ in the entire redshift range, which suggests that $\Lambda$CDM has a higher growth rate for mid-sized voids than larger voids, consistent with a lower merger rate than the other two models. 

Thirdly, we found that the preference for slightly elliptical voids over spherical ones at low redshifts is also present at higher redshifts. Smaller voids which are less likely to have undergone many mergers tend to exhibit more extreme ellipticities, pointing to the anisotropic, asymmetric nature of void growth (expansion of void walls). Larger voids tend to be closer to the average shape, though not entirely spherical. We proposed that this was due to the averaging out of more extreme shapes by merging. We also suggested that the commonality of slightly elliptical and prolate voids, no matter their volume, is due to anisotropic matter evacuation. These observations are the same across all three models. The ellipticity and prolateness distributions are indistinguishable between the models, and show no obvious traces of the underlying cosmological model. From this, we concluded that the shape of a void, and its evolution with redshift, is dominated by gravitational effects rather than cosmological effects, and hence shape distributions would not serve as a good probe of a dynamical scalar field or coupling. 

Finally, we discovered that the form of dark energy, be it a scalar field or the cosmological constant, and dark sector coupling, leaves an imprint on the average void density distributions. We found that the $\phi$CDM model produces emptier voids on average than CDE or $\Lambda$CDM across $z=1.5-12$, extending the results found in~\cite{2017MNRAS.468.3381A} for $z=0-1$. Our results support the proposal made by~\cite{2017MNRAS.468.3381A}, that the dynamical scalar field acts to evacuate voids faster than they otherwise would, and the coupling between dark matter and dark energy delays or slows this down due to the drag force acting on baryonic particles moving out of voids. Since this effect is predicted over a large range of redshifts ($z=0-12$), this signature is very promising as an observational probe of the dark sector, particularly as a probe of dynamical scalar fields. Additionally, as there is good agreement between other void finders and Hessian-based void finders, especially in the void density profiles, we expect this signature to be detectable through other popular void finders that have already been applied to observational data (e.g. \textsc{zobov}, \textsc{vide}). 

In summary, we have found a number of differences in void population properties between the three models. Specifically, our results suggest that the primary cause of all the discrepancies discovered is an increased evacuation rate from voids in the models containing a dynamical scalar field, which results in a greater merger rate for these models. While some of the imprints we have isolated in this study are not observable unless the void population studied is very large, and contains a sufficient number of large voids, we have found one very promising probe of the dynamical scalar field form of dark energy in the average density distribution of voids, across a wide range of redshifts. It is worth noting that although our proposed explanations are consistent with all of our results, without deeper analyses it is unclear if there are other effects at play we have not been able to isolate in this study (e.g. void wall expansion rate). In future studies, we will attempt to elucidate this situation by calculating void growth rate, wall expansion rate, merger rate and particle evacuation rate from voids for each model. While this signature is easily observable in a well-characterised cold dark matter density field, tracer bias may affect its visibility in an observational context, particularly at high redshift. For this reason, we will also consider the impact of using the galaxy distribution to derive the density field, and observational selection for galaxies at $z\approx 0-6$ and for 21 cm maps of the cosmic web at higher redshifts, on these signatures. Additionally, we will be investigating the usage of related and more robust observables to amplify the signal we found, in order to minimise the effect of observational limitations.  

With a more in-depth understanding of how alternative dark sector physics affects different processes of void evolution, and how observational biases affect signatures, we can develop promising probes of the cosmology of our Universe, and inform the way future large surveys are conducted so that we can effectively constrain the properties and the nature of the dark sector. 

\section*{Acknowledgements}
E. A. acknowledges financial support by the Australian Government and The University of Sydney, through an Australian Postgraduate Award and a University of Sydney Merit Scholarship, respectively. P. J. E. acknowledges funding from the SSimPL programme and the Sydney Institute for Astronomy (SIfA), DP130100117 and DP140100198. The authors acknowledge the University of Sydney HPC service at The University of Sydney for providing HPC resources, in particular the Artemis supercomputer, that have contributed to the research results reported within this paper. URL: \url{http://sydney.edu.au/research_support/}. This research made use of the NCI National Facility in Canberra, Australia, which is supported by the Australian Commonwealth Government, with resources provided by Intersect Australia Ltd and the Partnership Allocation Scheme of the Pawsey Supercomputing Centre. We thank Vasiliy Demchenko for enlightening discussions about our void finding method, and his suggestion to compare the average densities of individual voids with the mean density of the simulation box, as a means of determining the proportion of voids that could be voids-in-clouds. 




\bibliographystyle{mnras}
\bibliography{Paper1} 







\bsp	
\label{lastpage}
\end{document}